\documentclass[sigconf]{acmart}

\usepackage{booktabs} 
\usepackage{definitions}
\usepackage{algorithm}
\usepackage[noend]{algpseudocode}
\usepackage{color}
\usepackage{float}
 \usepackage{subfig}
 \usepackage{enumitem}

\setcopyright{rightsretained}

\acmConference[KDD'17]{ACM Woodstock conference}{August 2017}{Halifax, Nova Scotia Canada} 
\acmYear{2017}
\copyrightyear{2017}

\acmPrice{15.00}

\begin{document}
\title{Recurrent Poisson Factorization for Temporal Recommendation }

\author{Seyed Abbas Hosseini,\\Keivan Alizadeh, \\Ali Khodadadi, Ali Arabzadeh }
\affiliation{%
  \institution{Sharif University of Technology}
  \city{Tehran} 
  \country{Iran} 
}
\email{a_hosseini,kalizadeh,khodadadi,arabzadeh@ce.sharif.edu}

\author{Mehrdad Farajtabar, Hongyuan Zha}
\affiliation{%
  \institution{Georgia Institute of Technology}
  \city{Atlanta} 
  \state{GA}
  \country{USA}}
\email{mehrdad@gatech.edu, zha@cc.gatech.edu}

\author{Hamid R. Rabiee}
\affiliation{
   \institution{Sharif University of Technology}
  \city{Tehran} 
  \country{Iran} 
  }
\email{ rabiee@sharif.edu}

\renewcommand{\shortauthors}{A. Hosseini et al.}

\begin{abstract}
Poisson factorization is a probabilistic model of users and items for recommendation systems, where the so-called \emph{implicit consumer data} is modeled by a factorized Poisson distribution. 
There are many variants of \emph{Poisson factorization} methods who show state-of-the-art performance on real-world recommendation tasks. 
However, most of them do not explicitly take into account the temporal behavior and the recurrent activities of users which is essential to recommend the right item to the right user at the right time.
In this paper,  we introduce  \emph{Recurrent Poisson Factorization} (RPF) framework that generalizes the classical PF methods by utilizing a Poisson process  for modeling the implicit feedback. 

RPF treats time as a natural constituent of the model and brings to the table a rich family of time-sensitive factorization models. 
To elaborate, we instantiate several variants of RPF who are capable of handling dynamic user preferences and item specification (DRPF), modeling the social-aspect of product adoption (SRPF), and capturing the consumption heterogeneity among users and items (HRPF). 
We also develop a variational algorithm for approximate posterior inference that scales up to massive data sets.
Furthermore, we demonstrate RPF's superior performance over many state-of-the-art methods on synthetic dataset, and large scale real-world datasets on  music streaming logs, and user-item interactions in M-Commerce platforms.

\end{abstract}

%
%


%
\keywords{Poisson factorization, Poisson Process, Temporal recommender system}

\maketitle

\section{Introduction}
Recommendations are the main drive for consumer purchase in online retailers and the most effective factor in user engagement in online services. 
However, delivering the right item at the right time to the right person is a challenging task due to many reasons.
Firstly, the user preference is not directly observable and must be learnt from \emph{implicit feedback}.
Moreover, users may have different behavioral patterns;
Some tend to explore and test new products while others use a limited number of products frequently. 
In addition to this heterogeneity, their interest and preference change over time and may exhibit seasonality.
On the other hand, the ubiquity and increasing presence of social networks in everyday lives transforms the purchasing process  or service adoption from a simple matching of user preference and item specification to a more complex one affected by the peer influence. 

Poisson Factorization (PF) is the gold standard framework for recommendation systems with implicit feedback~\cite{HPF,SPF,DPF,NPPF,ContentBasedPF}.
The PF based methods utilize the user interactions with items in terms of the count of item usages, where
this implicit feedback signal helps infer the latent user preference and item specification.
There is a wealth of research to adapt the PF framework to tackle the aforementioned problems.
Importantly, diversity and heterogeneity of users and products is taken care of by an innovative hierarchical probabilistic model \cite{HPF} called HPF hereafter.
Dynamic Poisson Factorization (DPF) captures the time evolving latent factors with a Kalman filter~\cite{DPF}.
Moreover, to incorporate social network information into a traditional factorization method, Social Poisson Factorization (SPF) enriched the preference-based recommendations~\cite{SPF}.

However, the existing variants of PF based recommendation do not explicitly take into account the \emph{temporal} behavior and the \emph{recurrent} activities of users.
The time-sensitive product adoption data carries a great deal of information on usage dynamics which is otherwise neglected when the user-item interaction is only reported by the aggregated number of times they happen~\cite{TimeSVD, TimeSensitive}.
Furthermore, it allows successful predictions of the returning time which not only allows a web company to keep track of the evolving user preferences, but also results in more clever marketing strategies and time-sensitive recommendation. After all, online retailers need not blindly advertise all the times and make the users indifferent. 
Traditionally, in PF based recommendation the state of each user is often assumed
to be binary -- either adopting a product or not,  or an aggregated count/rating variable.
However, such assumption does not capture the recurrent nature of product usage, where the frequency and time of the usage matters.
It's noteworthy that the classical time-window or instance-decay approaches and manually discretization do not work, as they lose too much information when discarding data instances or coarse-grained discretizing of the time. 
Alternatively, one can improve the accuracy by more granular approximation at the cost of  increased computational complexity. 

In this paper, we introduce \emph{Recurrent Poisson Factorization} (RPF) for recommendation systems.
 RPF is a mathematical framework based on \emph{Poisson processes}~\cite{kingman1993poisson}, which generalizes the classical PF methods by modeling the product adoption as a non-homogeneous Poisson process over time rather than a Poisson distribution on the aggregated count values. 
 RPF jointly models the time and the type of items with which the user interacts and hence is able not only to predict the item but the time that  she will adopt a service.
A distinctive feature of RPF is its interpretability and intuitive parameterization;
In addition to following the classic PF methods on learning a latent representation for each user and item it  provides a sound and rigorous framework to model the effect of previous actions of the user and her friends on future service usages and preferences. 
RPF framework not only extends the basic PF method~\cite{HPF}, but is applicable to the many of its variants. 
Basically, we establish the foundation for replacing Poisson distribution with the Poisson process whenever there is a call for temporal and recurrent service adoption.
 Notably, we extend the well-known HPF, DPF, and SPF methods to their recurrent counterparts,  HRPF, DRPF, and SRPF, respectively. 
These downstream variants of RPF accompanied with its well-founded mathematical essence makes it a compelling replacement for classic PF when the temporal granularity and frequency of actions matter.
In summary, we have the following main contributions:
\begin{itemize}
	\item We establish a previously unexplored connection between Poisson factorization framework and Poisson processes, which allows to recommend the right item to the right user at the right time by inferring the dynamic user interests over time and predicting the time that users need each item.
	\item We show that RPF framework is applicable to many variants of basic PF. It can handle dynamic user preference and item specification (DRPF), It can capture the heterogeneity and diversity among users and items (HRPF), and It  is able to incorporate the peer influence and the network of users (SRPF).
	\item We propose a scalable variational inference algorithm based on mean-field approximation for RPF that scales up to millions of user-item interactions.
	\item We conduct several experiments on synthetic and two real datasets to demonstrate the performance of our model. To this end, we curated a dataset consisting of the listening history of more than 1000 users over 6 months including more than 450K user-song pairs.\footnote{The codes and data are available at \href{https://github.com/AHosseini/RPF}{https://github.com/AHosseini/RPF}}
\end{itemize}

{\bf Related Work}. 
The works most closely related to ours are roughly divided into two parts: Poisson factorization methods and Poisson point processes.  These two lines of research for recommendation systems are progressing independently, and, to the best of our knowledge, we are the very first to systematically combine these two and propose a unified framework for Poisson process factorization. 

Matrix factorization algorithms are widely used  for recommendation in order to infer the user preferences based on similar consumption patterns among users~\cite{NonNegativeMatrixFactorization}. Traditional Factorization methods used user ratings to products in order to estimate their preferences. Since users rarely provide such explicit feedbacks, more recent methods such as Poisson factorization infer such preferences from user implicit feedback. Different variants of PF are able to consider the heterogeneity among users, dynamic user interests over time and peer influence among users~\cite{HPF,DPF,SPF}. Moreover, the nonparametric version of PF is able to effectively estimate the dimension of latent feature space~\cite{NPPF}. Collaborative Topic Poisson Factorization (CTPF) is another variant of PF which is customized for article recommendation and is able to consider the content of articles to infer the user preferences more effectively~\cite{ContentBasedPF}. Although PF provides a general framework for  recommendation with many desirable features, but looses a great deal of information due to ignoring the temporal patterns of user actions.

More recently, point process methods have attracted a lot of interest in user behavior modeling~\cite{khodadadi2017umub} and recommendation systems~\cite{TimeSensitive,NeuralSurvival,RecurrentMarkedTPP,CoevolutionaryRecommenderSystem}.
The paper \cite{CompetingProducts} presents a first Hawkes process approach to model the adoption of products, however, their model does not capture dynamic preferences and suffer from inconsistent  and negative rate functions. The method proposed in \cite{C4} models the adoption of competing products by a multidimensional point process. However, it doesn't provide a method for recommending appropriate items to users.
The authors in~\cite{TimeSensitive}  propose a compelling point process model for recommendation systems, however, the model lacks the peer influence within the network and suffers from non-dynamic user preference and item specifications. Moreover, to model the heterogeneity between users they impose a low-rank regularizer which blindly set most latent dimensions of preferences to zero while keeping a few non-zero elements. In contrast, our approach captures the diversity between users by hierarchal Bayesian modeling.  \cite{CoevolutionaryRecommenderSystem} proposed a coevolutionary model of user and item latent features without incorporating network influence or considering the heterogeneity among users.  Last but not least, all these work lack a systematic treatment of Poisson factorization methods and Poisson processes and their close connections. 

\begin{figure*}[!t]\label{fig:RPFIntuition}
	\includegraphics[width=5in]{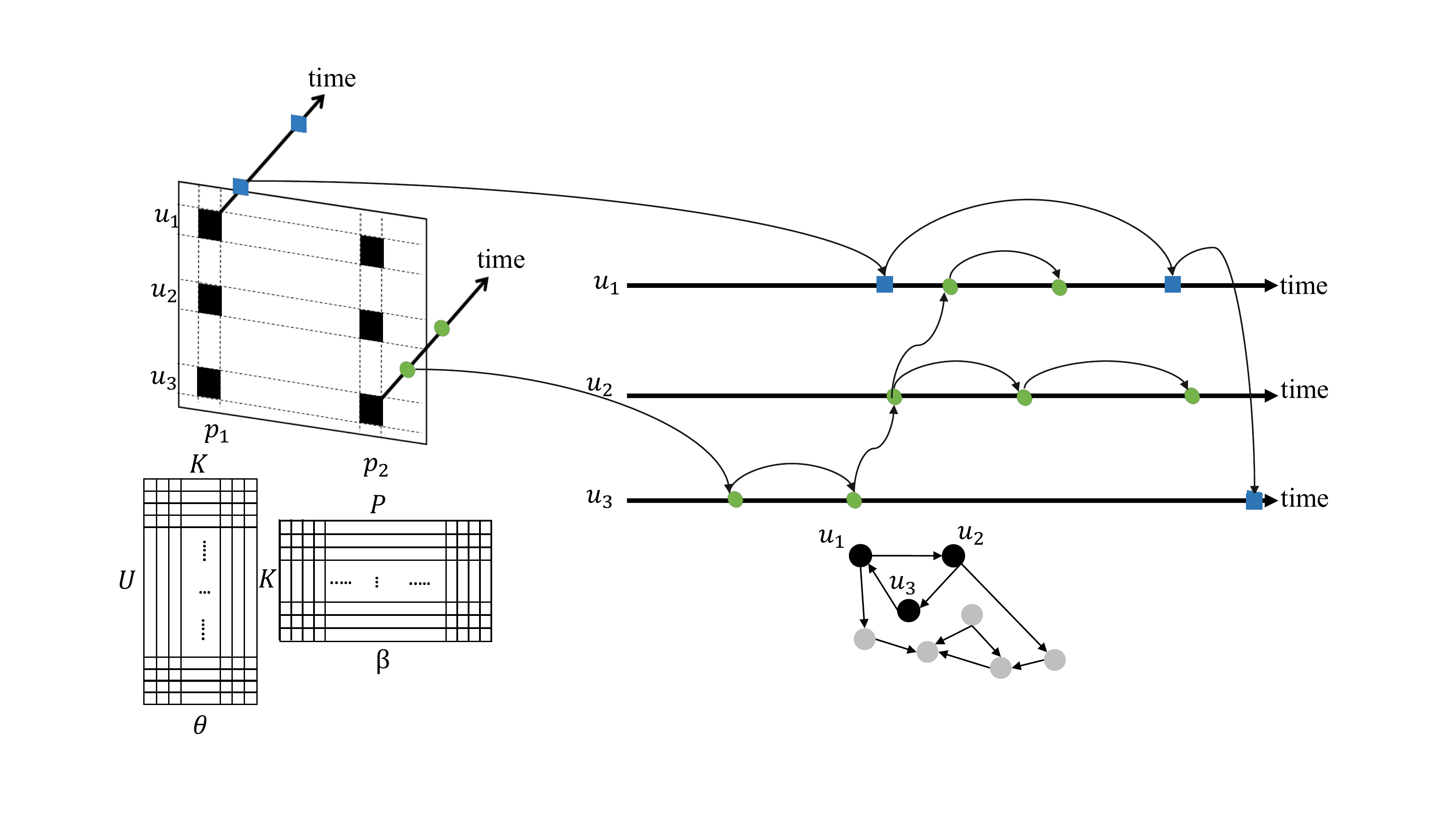}
	\vspace{-3mm}
	  \caption{The illustration of the RPF model. The left matrix represents the user-item engagement that are triggered by user intrinsic interest on different items and are inferred using the factorization. Each of these events may result in future engagement of the same user or one of her friends with the same item. These events and the triggering relations among them are represented in the right plot. The arrows show the triggering relationship among events. The social relations are also depicted in the right bottom. The arrows show the following relationship among users.}\label{fig:RPFIntuition}
\end{figure*}

\section{Preliminaries}\label{sec:Preliminaries}

\subsection{Poisson Factorization}
Poisson factorization is a general framework for recommendations based on users implicit feedbacks, such as clicks, purchases or watches. PF methods use the number of times a user $u$ consumed the product $p$ to infer the interest of user to all products including non-consumed ones. HPF is the most basic model in this framework, which assumes  the number of times that user $u$ purchased or clicked an item $p$ is a Poisson random variable $r_{up}$ with rate $\lambda_{up}$. HPF models the  interest of user $u$ and properties of product $p$ as $K$ dimensional latent vectors $\theta_u$ and $\beta_p$, respectively, and use their inner product as the latent rate of $r_{up}$, that is:
\begin{align}\label{eqn:HPF}
	r_{up}\sim \mathbf{Poisson}(\theta_u^{\top}\beta_p).
\end{align}
In order to model the heterogeneous interests of users and  the variety of items, HPF uses a hierarchical prior over latent feature vectors \cite{HPF}. Although this model has rigorous  statistical properties such as sparse representation and ability to capture the long-tailed user activities, unfortunately, it is  not able to utilize the social relation among users that is inherent to many recommendation scenarios.

Social Poisson Factorization (SPF) is another development in PF framework that solves the above issue \cite{SPF}. SPF models the number of times user $u$ is engaged with an item $p$ is a Poisson random variable. However SPF assumes that  the rate of $r_{up}$ is the sum of $\theta_u^{\top}\beta_p$ and a weighted sum of the number of times user $u$'s friends have consumed this item. That is:
\begin{align}\label{eqn:SPF}
	r_{up}|r_{-up}\sim \mathbf{Poisson}(\theta_u^{\top}\beta_p+\sum_{v\in N(u)}\tau_{vu}r_{vp}),
\end{align}
where $\tau_{vu}$ is the latent impact of user $v$ on user $u$. In this way, SPF  incorporates the social relations among users with the user-item history in an effective way. Therefore, it's able to infer the influence of users on each other. However, since this model is not able to consider time of users' actions,  the rate of adopting item $p$ by user $u$  will be increased by $\tau_{vu}$ any time after her friend $v$ adopts this item. This is not true in real-world scenarios as the effect of users' actions on their friends usually decays after some time. 

The PF factorization framework has advanced continuously to handle more real-world and applied situations. The  evolving user preferences and item specification (over time) was one of the serious challenges of recommender systems. In appeal to this call, Dynamic Poisson Factorization (DPF) is introduced as a recommendation method based on Poisson factorization. It basically solves this issue by considering time dependent feature vectors for users and items. DPF is a discrete-time approach which models the  evolution of users and items latent features over time by a Kalman filter. In order to do so, DPF models latent features  by a Gaussian state space model and exponentiates the state space model to make it nonnegative. Therefore, the rate of user $u$ for adopting item $p$ becomes:
\begin{align}\label{eqn:DPF}
	r_{up,t}\sim \mathbf{Poisson}(\sum_{k=1}^Ke^{(\theta_{ukt}+\bar{\theta}_{uk})}e^{(\beta_{pkt}+\bar{\beta}_{pk})})
\end{align}

One of the most limiting properties of the PF methods is that they summarize the history of user-item interactions in terms of a single count and hence discard a great deal of information. Furthermore, their are not usually well-suited to model the situations where the consumption is a recurrent process. Failing to answer time-sensitive questions is another drawback of the classic PF models. 
In the next section, we make a brief introduction to Poisson processes which are able to capture the temporal dynamics of user-item interactions more effectively. 
\subsection{Poisson Process}
Poisson process is a stochastic process which is suitable for modeling timestamped events. This process assumes that the number of events in a time window $[t_0,t_1]$, $N_{[t_0,t_1]}$, is a random variable distributed as:
\begin{align}
	N_{[t_0,t_1]}\sim \mathbf{Poisson}(\lambda\times (t_1-t_0)),
\end{align}
where $\lambda$ is the rate of occurring events. The main advantage of Poisson process over Poisson distribution in modeling timestamped events is the fact that it considers the interval in which the events occur and hence can  estimate the rate of occurring events more accurately. To elaborate, consider two scenarios: in the first one, three events occur during one month  \textit{e.g.} at days 1, 10, 25.   Now assume in the second scenario all the three events occur in a single day, say  day 20. The PF based models consider both scenarios the same because Poisson distribution can only work with the aggregated counts. In contrast poisson processes can differentiate between the two scenarios: The first process has higher rate of occurrence in the beginning of the months, while the second process has high activity rates towards the end of the month.

In real-world applications events and times series data are not usually distributed uniformly over the time. For example, the distribution of inter-event durations in many applications such as posting emails, listening to music, twitting and checking into social networks have been shown to be bursty, heavy-tailed and periodic \cite{STP}. Hence they can't be effectively modeled using poisson distribution and  even homogeneous Poisson processes. \emph{Non-homogeneous Poisson process} is a process with dynamic rate of event occurrence that can capture the complex longitudinal dependencies among events. This process can be uniquely defined using the so-called \emph{conditional intensity function}, $\lambda^*(t)$. It encodes the expected rate of occurring event at time $t$ given the history of past events, $\mathcal{H}(t)$, \emph{i.e.},
\begin{align}\label{eqn:lambda}
\lambda^*(t)dt = \mathbb{P}\{\text{event in} [t,t+dt)|\mathcal{H}_t\}.
\end{align}
 Using the definition of $\lambda^*(t)$ in eq. \eqref{eqn:lambda}, the likelihood a list of events ($t_1, \ldots, t_n$) observed during  a time window $[0,T)$, can be written as 
\begin{align}\label{eqn:PoissonLikelihood}
\mathcal{L} (t_1, \ldots, t_n)= \prod_{i=1}^{n} \lambda^*(t_i)\underbrace{ \exp \left(-\int_0^T\lambda^*(s) ds \right)}_{\mathcal{S}(T)}.
\end{align}
Roughly speaking, the first term denotes the likelihood of occurring event at times $(t_1, \ldots, t_n)$, and the second term $S(T)$ denotes the likelihood of not occurring events elsewhere. Therefore, if the temporal pattern of occurring events is not uniform the likelihood is maximized  by an intensity function that describes the temporal dynamics the best. 

One of the most important properties of Poisson processes which makes them an appropriate candidate for modeling temporal events is the fact that the superposition of independent Poisson processes is itself a Poisson process with intensity equal to the summation of other intensities~\cite{kingman1993poisson}. We use this property to propose a computationally appealing framework for modeling user preferences over time, predicting their future actions, and timely recommending them the appropriate services and products.

\section{Recurrent Poisson Factorization}
\label{sec:proposed}
Intuitively, engagement of users with different items is driven by two main factors. Intrinsic user preferences and previous pleasant experiences of her friends or her own. For example, a user in last.fm may listen to an album due to her interest to the genre of the album or she may see her friends listening to the album and hence play it. Figure \ref{fig:RPFIntuition} shows such behavioral pattern. The left matrix shows that user $u_1$ and $u_3$ have listened to music $p_1$ and $p_2$ respectively because of their own initiative. The social network among the users is plotted in the bottom right. The upper right plot shows the events that are triggered by previous events and the arrows show the triggering relation among them. Since user $u_2$ follows user $u_3$ and user $u_3$ listens to music $p_2$, after some time, user $u_2$ also listens to that music. Furthermore, interest in this song may become viral and $u_1$ also plays it under influence from her followee, user $u_2$.

Recurrent Poisson Factorization (RPF) is a mathematical framework which models such behavioral patterns and is able to recommend the right item to the right user at the right time by utilizing the recurring temporal patterns in user-item engagement. In the followings, we first specify some notations and then discuss the generative model of RPF.

\subsection{Notations and Conventions}
Let $\mathcal{H}(T) = \{e_i\}_{i=1}^{M(T)}$  denotes the set of user-item interactions until time $T$ where $M(T)$ is the number of interactions up to time $T$.  The interaction $e_i$ is a triple $(t_i,u_i,p_i)$ which indicates that at time $t_i$, user $u_i$ interacted with product $p_i$. In addition to user-item interaction data, we may have the users social relations. In that case, a user $u$ may follow a set of  users which is denoted by $N_u$. For clarity, we use the following notation for the remainder of this paper. $\mathcal{H}_{up}(t)$ denotes the set of interactions of user $u$ with item $p$ until time $t$. We use dot notation to represent union over the dotted variable, \emph{e.g.}, $\mathcal{H}_{u .}(t)$ represents the events of user $u$, before time $t$, with any product, and $\mathcal{H}_{\bar{u}k}(t)$ denotes the interactions of all users except $u$, before time $t$, with item $i$.
\subsection{Proposed Generative Model}
The main idea of RPF is modeling the time of engagement of users with different products as a set of dependent Poisson processes where the intensity of each process consists of two main parts. The intrinsic intensity which denotes the matching between user preferences and items attributes, and the extrinsic intensity which represents the tendency of user to engage with an item triggered by previous user-item interactions. RPF assumes a  hierarchical matrix factorization model  for intrinsic users interests on items and models the triggering effect of engagement of a user with an item on future user-item interactions using a set of cascade Poisson processes\cite{JordanPoissonProcess}. In other words, it is 
assumed that the times of interacting user $u$ with item $p$ is generated by a Poisson process with the following intensity:
 \begin{align}
     \lambda_{up}(t) = \theta_u^{\top}\beta_p+\sum_{e\in H(t)}\kappa_{up}(e,t),
 \end{align}
where $\theta_u$ and $\beta_p$ are the $K$ dimensional latent vectors which denote the interests of user $u$ and attributes of item $p$, respectively. The dot product  between user interests and the item attributes essentially shows the similarity  between them  and serves as the user intrinsic intensity to engage with this item. 
In order to model the effect of previous user-item interactions on the tendency of user $u$ to adopt item $p$ at time $t$, it is assumed that each of the previous user-item interactions $e$, triggers a Poisson process which increases the tendency of user $u$ to engage with item $p$ at time $t$ by $\kappa_{up}(e,t)$. Since the effect of  this interaction usually decreases over time, the exponential function is widely used. That is, in the simplest form, the kernel $\kappa_{up}(e,t)$ can be defined as:
\begin{align}
    \kappa_{up}(e,t) = e^{-\omega(t-t_e)}\mathcal{I}(p_e=p,u_e=u),
\end{align}
where, $\mathcal{I}$ is the indicator function which shows that the tendency of  user $u$ to adopt item $p$ only depends on her own experiences with that item.  $\kappa_{up}(e,t)$ can be any integrable function of time and can take a variety of functional forms to capture the complex temporal properties of preference propagation among users and items.

As it was mentioned before, RPF is a general framework for recommendation  which can be customized to effectively include different types of information and priors and beliefs. We will introduce three variants of this model in the following.

{\bf Hierarchical Recurrent Poisson Factorization (HRPF)} assumes a hierarchical prior over the set of $\{\theta_u\}_{u=1}^U$ and $\{\beta_p\}_{p=1}^P$  in order to promote and model the diversity of users and items. Since the gamma distribution is conjugate with the likelihood in eq. \eqref{eqn:PoissonLikelihood} and also constrains the latent vectors to be nonnegative and sparse, we use the same idea as in \cite{HPF} and consider a gamma prior over these variables with a latent rate parameter, \emph{i.e.}
\begin{align}
    \theta_u\sim\text{Gamma}(a^{\theta,\text{shp}},\eta_u )\\\nonumber
    \beta_p\sim\text{Gamma}(a^{\beta,\text{shp}},\xi_p ),
\end{align}
where, $\eta_u$ and $\xi_p $ are gamma distributed random variables equip RPF with a heterogeneous users and items latent features.

{\bf Social Recurrent Poisson Factorization} is a variant of RPF that is able to model the impact of social network on users' actions. In this model, we assume  the actions of user $u$ are either triggered by her intrinsic interests or by previous actions of her friends. To this end, we propose the following intensity function for user $u$'s interaction with item $p$:
\begin{align}
    \lambda_{up}(t) = \theta_u^{\top}\beta_p+\sum_{v\in \mathcal{N}_u}\sum_{e\in H_{vp}(t)} \tau_{vu}g_{\omega}(t_i,t),
\end{align}
where, $\tau_{vu}$ is the latent influence of user $v$ on user $u$ and $g_{\omega}(t_i,t)$ is the temporal kernel with parameter $\omega$  
capturing the complex longitudinal dependencies among events. 
This kernel can be any integrable function that represents the impact of an event at time $t_i$ on another event at time $t$. In order to make the influence matrix $\tau$ nonnegative and sparse, we assume a  gamma prior over  these variables which is also conjugate with likelihood in eq. \eqref{eqn:PoissonLikelihood}.

{\bf Dynamic Recurrent Poisson Factorization (DRPF)} is another variant of RPF which models the dynamic interests of users and popularity of items over time. DRPF proposes the following intensity for engaging user $u$ with product $p$:
\begin{align}
    \lambda_{up}(t) = \theta_u^{\top}(t)\beta_p(t)+\sum_{e\in H_{up}(t)} \tau_{uu}g_{\omega}(t_i,t),
\end{align}
where, $\theta_u(t)$ and $\beta_p(t)$ are two processes which denote the preferences of user $u$ and attributes of item $p$ at time $t$, respectively.  $\tau_{uu}$ is a latent parameter which shows the amount of influence of user $u$ by her own previous interactions with items. In order to parametrize $\theta_u(t)$ and $\beta_p(t)$, we assume that these are nonnegative linear combination of simpler processes:
\begin{align}
        \theta_u(t) = \sum_{i=1}^I{\theta_u^ih_i(t)}\\
        \beta_p(t) = \sum_{j=1}^{J}{\beta_p^jl_j(t)}
\end{align}
where $h_i(t)$ and $l_j(t)$ are known functions over time and  $\theta_u^i$ are nonnegative latent variables that should be learnt from data. For example, since the users have similar activity patterns in similar days of week or similar hours of the day, we can use  24-dimensions function $h(t)$,  which $h_i(t)$ is $1$ only in the $i$th hour of the day. Furthermore, an extra 7 dimensions  for weekdays is also added:
\begin{align}
    h_i(t) = \begin{cases}
        \mathcal{I}(\text{hour(t) = i  }) \qquad  & 1\leq i \leq 24 \\
        \mathcal{I}(\text{day(t) = i-24  }) \qquad  & 25 \leq i \leq 31 \\
    \end{cases}
\end{align}
and hence we can learn the preference of users in different hours of  the day and different days of the week.

{\bf Dynamic Social Recurrent Poisson Factorization (DSRPF)} is a variant of RPF which combines the features of SRPF and DRPF to jointly models the dynamic interests of users and popularity of items over time and peer influence among users in social network. DSRPF use the following intensity function for the user-item interaction:
\begin{align}
    \lambda_{up}(t) = \theta_u^{\top}\beta_p+\sum_{v\in \mathcal{N}_u}\sum_{e\in H_{vp}(t)} \tau_{vu}g_{\omega}(t_i,t)
\end{align}
\subsection{Prediction and Recommendation using RPF}\label{subsec:Recommendation}
RPF models the tendency of users to engage with different items over time using the proposed intensity function and hence is able to recommend the most appropriate item to the users at any time. In order to do so, we should infer the posterior of the latent variables of the model and compute the expected intensity of user to engage with different items and recommend the items to the user sorted based on the following expected intensity:
\begin{align}
    \mathbb{E}[\lambda_{up}(t)]=  \mathbb{E}[\theta_u]^{\top}\mathbb{E}[\beta_p]+\sum_{v\in \mathcal{N}_u}\sum_{e\in H_{vp}(t)} \mathbb{E}[\tau_{vu}]g_{\omega}(t_i,t)
\end{align}
Moreover, RPF models are able to predict the returning time of the users to products. To this end we adopt the approach introduced in \cite{TimeSensitive}, i.e. we sample the time of next event from the Poisson process with intensity $\mathbb{E}[\lambda_{up}(t)]$ using Ogata's thinning algorithm \cite{ogata1981lewis} and report the sample mean as the expected returning time of user to the item.

\section{Inference}
As it was mentioned in section \ref{subsec:Recommendation}, the performance of the model to recommend the right item to the users and predicting the returning time of the users accurately, heavily depends on inferring the posterior of the latent variables efficiently. Since DSRPF model contains all features of other variants of RPF, in this section we propose a scalable inference algorithm for DS-RPF-based on mean-field variational inference \cite{jordan1998learning}. 

In order to make the model conditionally conjugate and obtain simple updates, we introduce an auxiliary variable $s_N$ for each user-item interaction $e_i$, which denotes the factor that triggered $e_N$. Since the triggering factor of a user-item interaction is either one of its previous events or matching between one of the user $u$'s feature vector component and product $p$'s features, therefore, we can define the conditional intensity of an event and its triggering factor as:
\begin{align}\label{eqn:completeLambda}
    \lambda^*_{up}(t,s)  = \begin{cases}
         \theta_{uk_s}^{i_s} \beta_{pk_s}^{j_s}h_i(t)l_j(t) \qquad &-K\times I \times J< s \leq 0 \\
         \tau_{u_su}g_\omega(t-t_s)\qquad &0< s < N
    \end{cases} 
\end{align}
Since it's not possible to find the exact posterior of the latent variables, we consider a factorized distribution over all latent variables and find the distribution that is most similar to the posterior using Bayesian mean field approximation. Namely,
\begin{align}
q(S, \beta, \theta, \tau, \xi, \eta, \mu) = \prod_{e\in \mathcal{H}}q(s_e|\gamma^s_e)\prod_{p,k,j} q(\beta_{pk}^j~|~\gamma_{pk}^{\beta,j})\\\nonumber
\prod_{u,k,i} q(\theta_{uk}^i~|~\gamma_{uk}^{\theta,i})
\prod_u q(\xi_u~|~\gamma_{u}^\xi)
\prod_p q(\eta_p~|~\gamma_{p}^\eta)\\\nonumber
\prod_{u,v} q(\tau_{uv}~|~\gamma_{uv}^\tau)\prod_{u}q(\mu_u~|~\gamma_{u}^{\mu}),
\end{align}
where, $q(s_e|\gamma^s_e)$ is a multinomial distribution and all other factors are gamma distributed. Using mean field theorem, we know that the optimal approximate functions are as follows:
\begin{align}\label{variationalUpdate}
    \ln q_j^*(\cdot) = \EE_{i\neq j} [\ln p(E, S,\beta, \theta, \tau,\eta,\xi,\mu)] + \mathrm{const}
\end{align}
Therefore, we iterate over all factors and update their parameters based on eq. \eqref{variationalUpdate}. The inference algorithm is described in Alg.~\ref{alg:RPF_inference}.
\alglanguage{pseudocode}
\begin{algorithm}[!        t]
\small
\caption{DSRPF Variational Inference}
\label{alg:RPF_inference}
\begin{algorithmic}[1]
\State Initialize $\gamma$ randomly
\While{$\Delta\ell > \epsilon$} \Comment{\mbox{check for model convergence}}
\For{$e_n\in D$}
\For{$m= -I\times J\times K:-1$}
\State $\gamma_{n}^{s}(m) \propto \exp(\EE[\ln (\theta_{u_nk_m}^{i_m})]+\EE[\ln (\beta_{p_nk_m}^{j_m})]) $
\EndFor
\For{$m= 1:n-1$}
\State $\gamma_{n}^{s}(m) \propto \exp(\EE[\ln (\tau_{u_mu_n})])g_{\omega}(t_m,t_n)$
\EndFor
\EndFor
\For{all users $u\in U$}
\State $\gamma_{u}^{\mu,shp} \leftarrow \lambda_{u}^{\mu,\text{shp}}+|\{v:u\in N(v)\}|\lambda_{uv}^{\tau,\text{shp}}$\Comment{\mbox{Update $q(\mu_u)$}}
    \State $\gamma_{u}^{\mu,rte}\leftarrow \lambda_{u}^{\mu,\text{rte}}+\sum_{v:u\in N(v)}\EE[\tau_{uv}]$
    \State $\gamma_{u}^{\eta,shp} \leftarrow \lambda_{u}^{\eta,\text{shp}}+KI\lambda_{u}^{\theta,\text{shp}}$ \Comment{\mbox{Update $q(\eta_u)$ }}
        \State $\gamma_{u}^{\eta,rte}\leftarrow \lambda_{u}^{\eta,\text{rte}}+\sum_{k=1}^{K}\sum_{i=1}^I\EE[\theta_{uk}^i]$
\For{$v\in N(u)$} \Comment{\mbox{Update $q(\tau_{uv})$}}
    \State $\gamma_{vu}^{\tau,\text{shp}} \leftarrow \lambda_{vu}^{\tau,\text{shp}} + \EE_S[c_{vu}]$ 
    \State $\gamma_{vu}^{\tau,\text{rte}} \leftarrow \sum_{e\in E_v}G_{\omega}(T-t_e) + \EE_{\mu_v}[\mu_v]$
\EndFor
\For{$k=1:K$}\Comment{\mbox{Update $q(\theta_{uk}^i)$}}
\For{$i=1:I$}
\State $\gamma_{uki}^{\theta,\text{shp}} \leftarrow \lambda_{u}^{\theta,shp}+\sum_{p \in P}\sum_{j=1}^J\EE[c_{upk}^{ij}] $ 
\State $\gamma_{uki}^{\theta,\text{rte}} \leftarrow \sum_{j\in J}F_{ij}(T)\sum_{p \in P}{\EE[\beta_{pk}^j]}+\EE[\eta_u]$
\EndFor
\EndFor
\EndFor
\For{all products $p\in P$}
\State $\gamma_{p}^{\xi,shp} \leftarrow \lambda_{u}^{\xi,\text{shp}}+KJ\lambda_{p}^{\beta,\text{shp}}$\Comment{\mbox{Update $q(\xi_p)$  parameters}}
    \State $\gamma_{p}^{\xi,rte}\leftarrow \lambda_{p}^{\xi,\text{rte}}+\sum_{k=1}^{K}\sum_{j=1}^J\EE[\beta_{pk}^j]$
\For{$k=1:K$}\Comment{\mbox{Update $q(\beta_{pk}^j)$}}
\For{$j=1:J$}
\State $\gamma_{pkj}^{\beta,\text{shp}} \leftarrow \lambda_{p}^{\beta,shp}+\sum_{u \in U}\sum_{i=1}^I\EE[c_{upk}^{ij}]$ \State $\gamma_{pkj}^{\beta,\text{rte}} \leftarrow \sum_{i\in I}F_{ij}(T)\sum_{u \in U}\EE[{\theta_{uk}^i}]+\EE[\xi_p]$
\EndFor
\EndFor
\EndFor
\EndWhile
\end{algorithmic}
\end{algorithm}
\section{Experiments}
We evaluate the performance of RPF on large-scale synthetic and real world datasets. We show that RPF  not only model the recurrent user behaviors over time, but also effectively captures the changes in trends and user interests where other temporal models are incapable of.

\begin{figure*}[t]
\centering
\subfloat[]{\includegraphics[height=1.3in,width=1.65in]{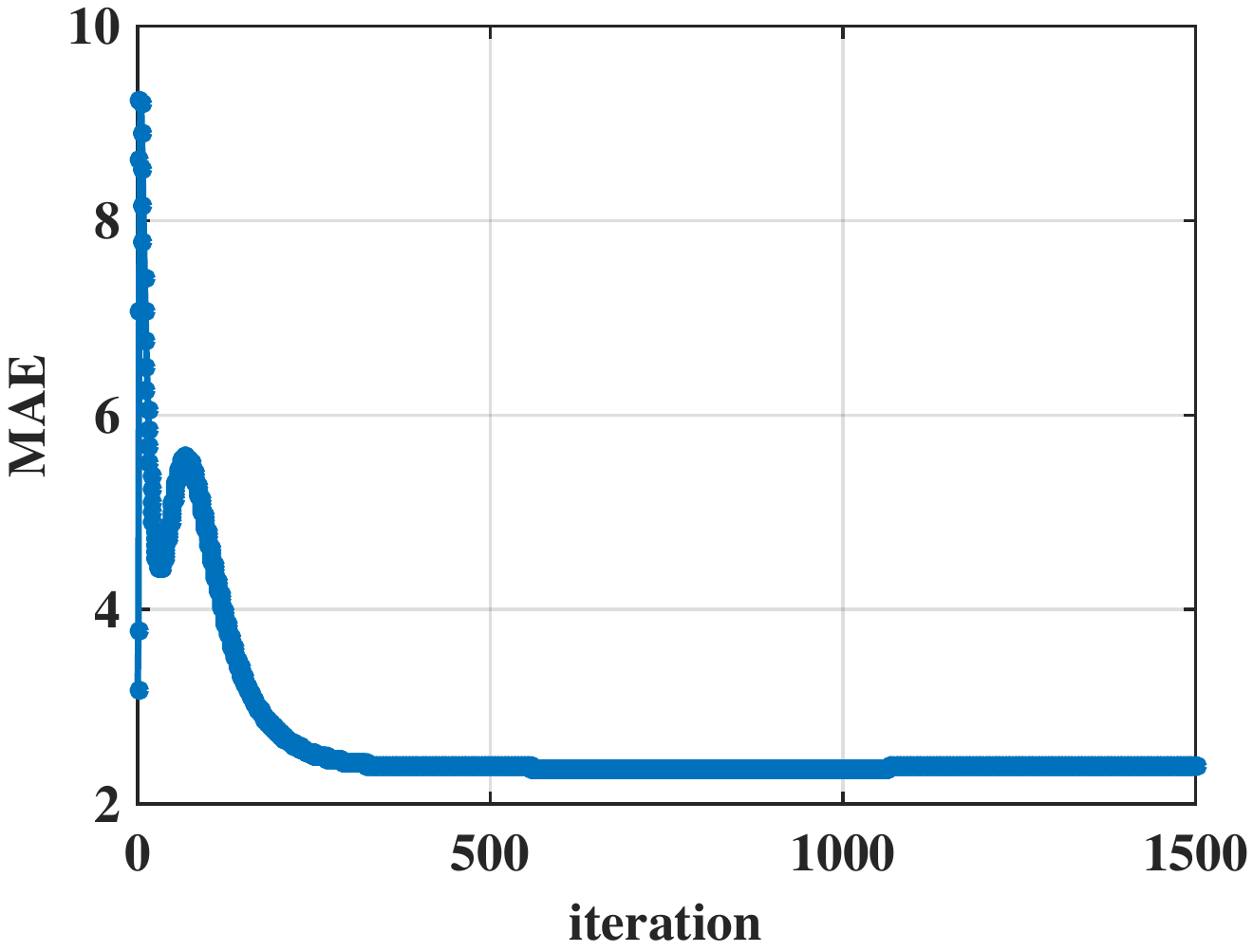}%
\label{fig:MAEOverIterations_Synth}}
\hspace{0.01in}%
\subfloat[]{\includegraphics[height=1.3in,width=1.65in]{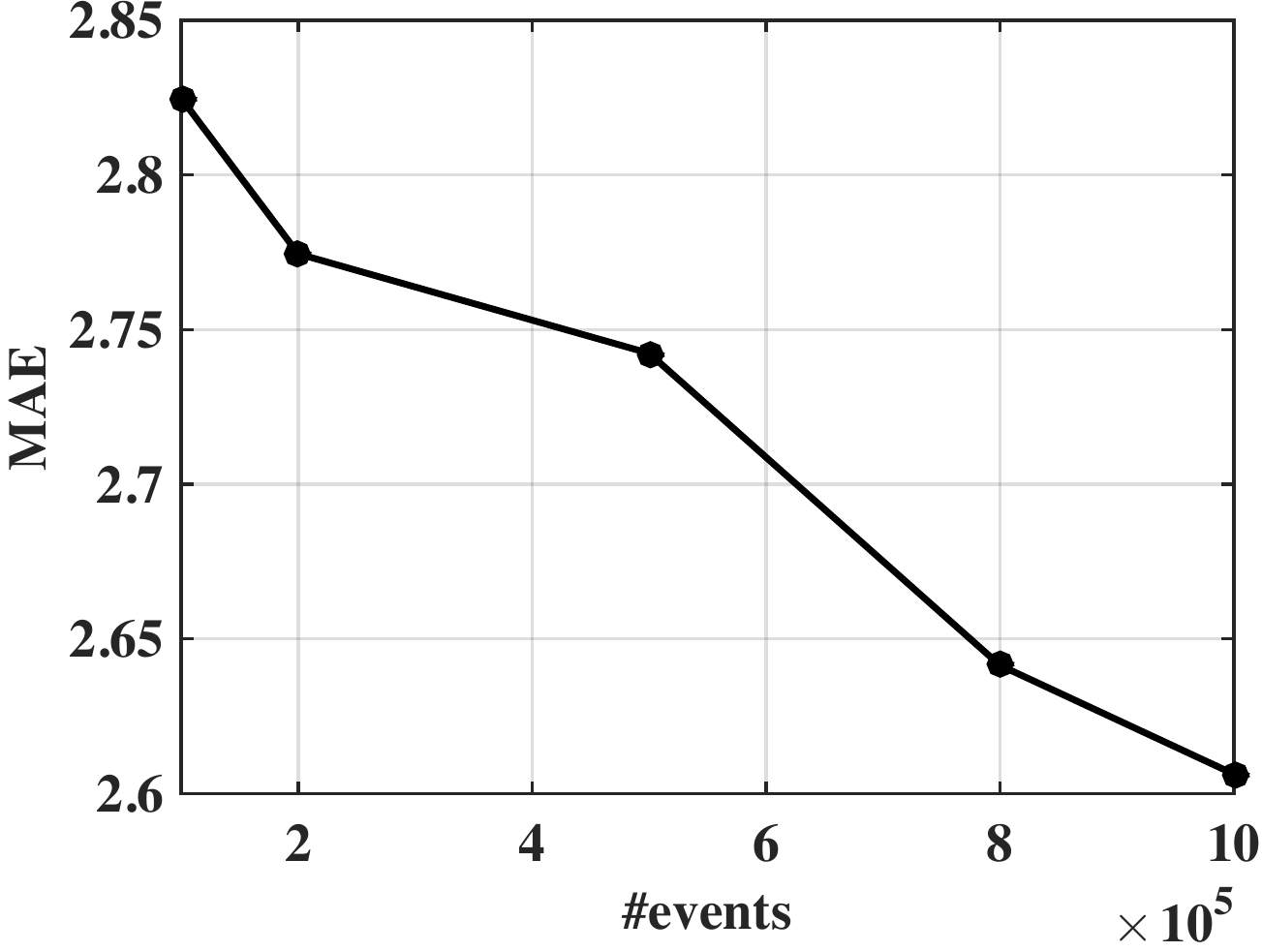}%
\label{fig:MAEOverEvents_Synth}}
\hspace{0.1in}%
\subfloat[]{\includegraphics[height=1.3in,width=1.65in]{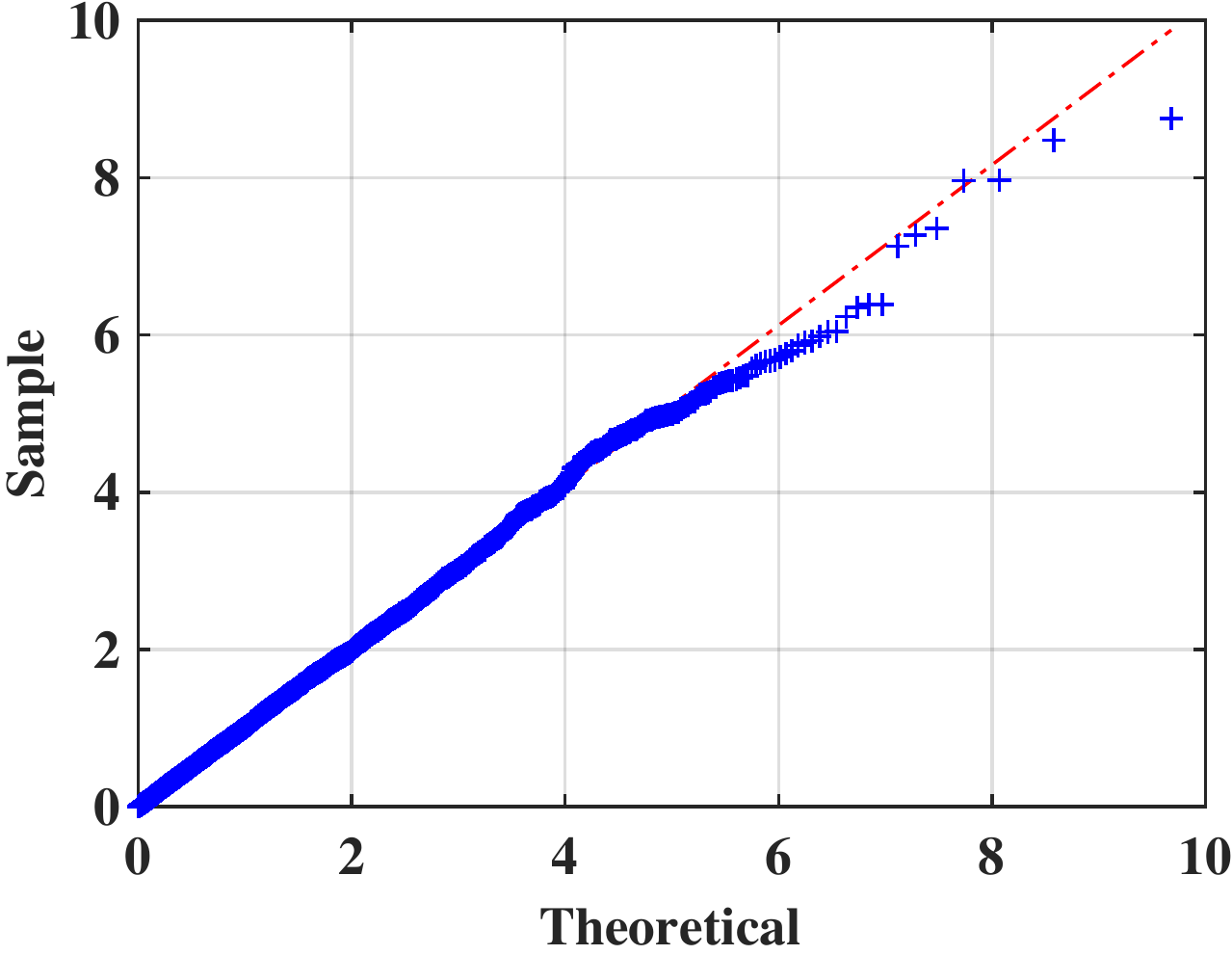}%
\label{fig:QQPlot_Synth}}
\hspace{0.1in}%
\subfloat[]{\includegraphics[height=1.35in,width=1.65in]{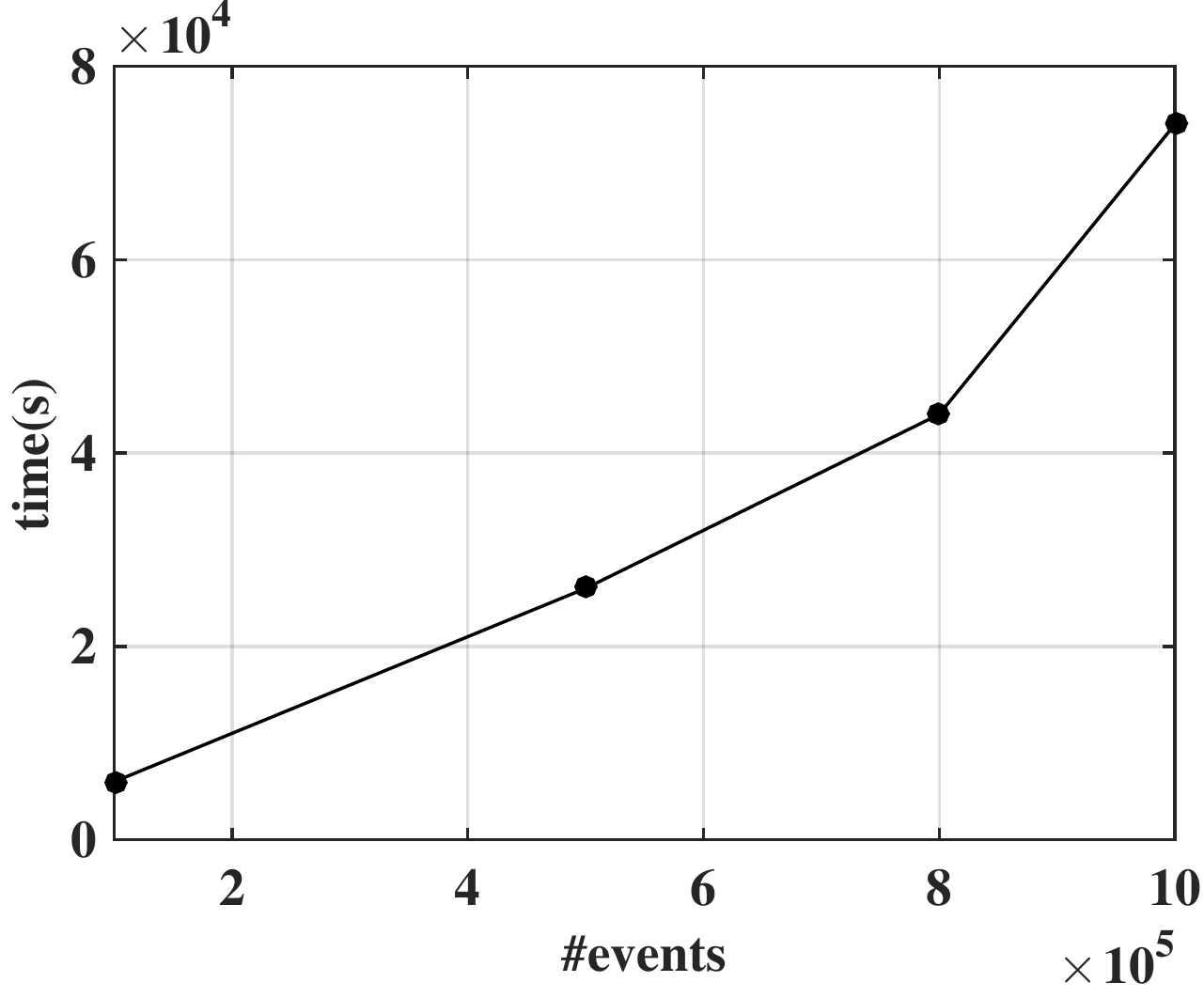}%
\label{fig:Scalability_Synth}}
\caption{The performance of proposed inference algorithm over synthetic data. (a) Mean Absolute Error (MAE) vs. the number of iterations, (b) MAE vs the number of events. 
(c) Quantile plot of intensity integrals, 
(d) Run-time vs. the number of events.
}
\label{fig:synt_performance}
\end{figure*}

\begin{figure*}[!t]
\centering
\makebox[10pt]{\raisebox{40pt}{\rotatebox[origin=c]{90}{\bf Last.fm}}}
\subfloat[]{\includegraphics[width=1.6in]{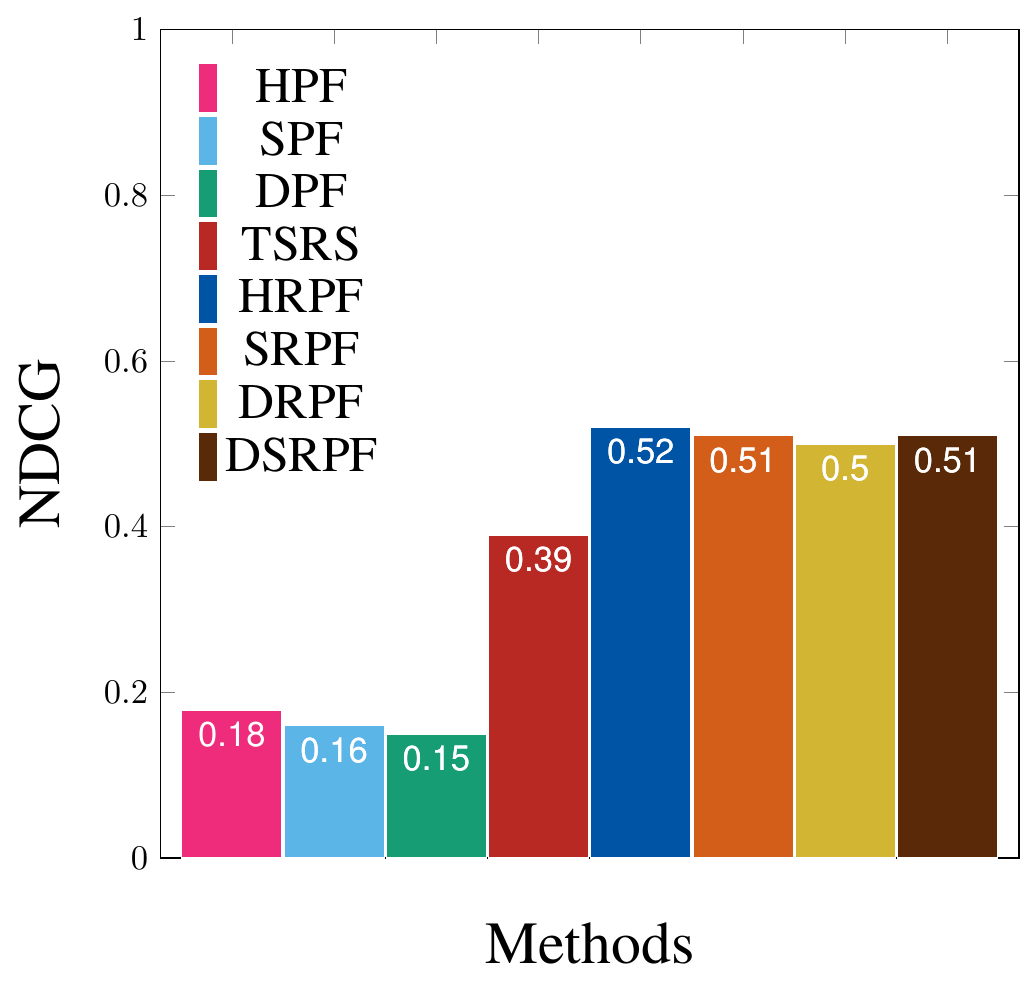}%
\label{fig:NDCG_LastFM}}
\hspace{0.01in}%
\subfloat[]{\includegraphics[width=1.6in]{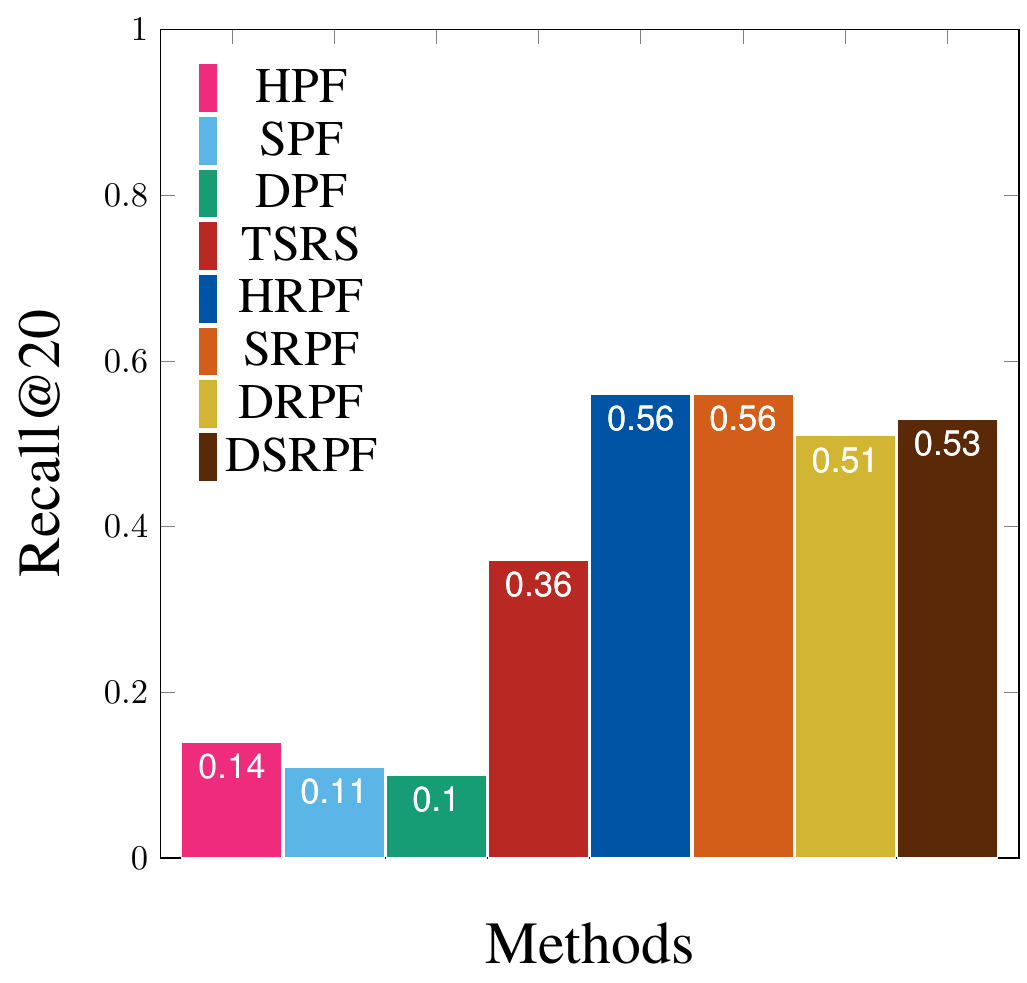}%
\label{fig:RecallAt20_LastFM}}
\hspace{0.1in}%
\subfloat[]{\includegraphics[height=1.5 in,width=1.6in]{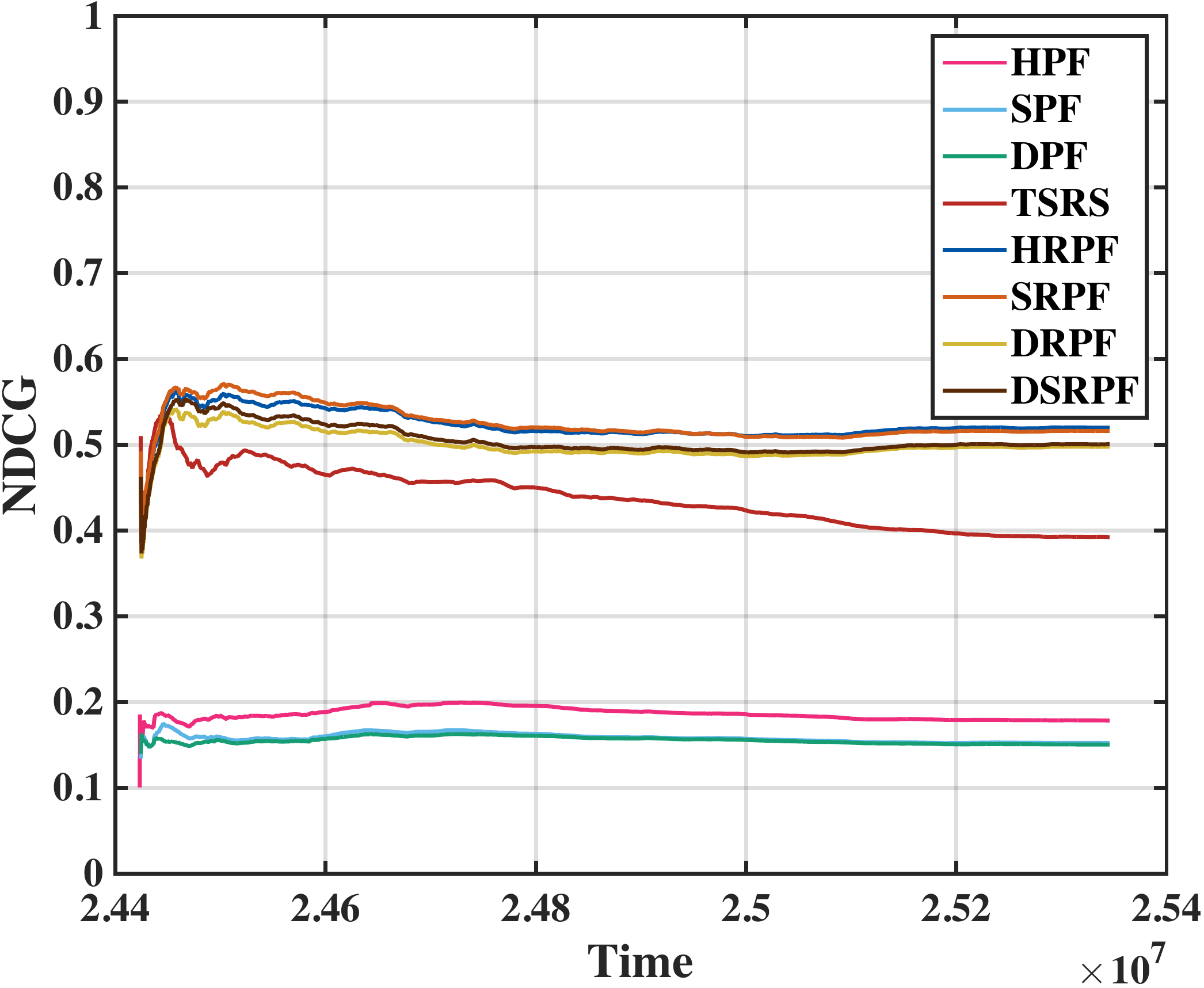}%
\label{fig:LastFM_over_time}}
\hspace{0.1in}%
\subfloat[]{\includegraphics[height=1.5in,width=1.6in]{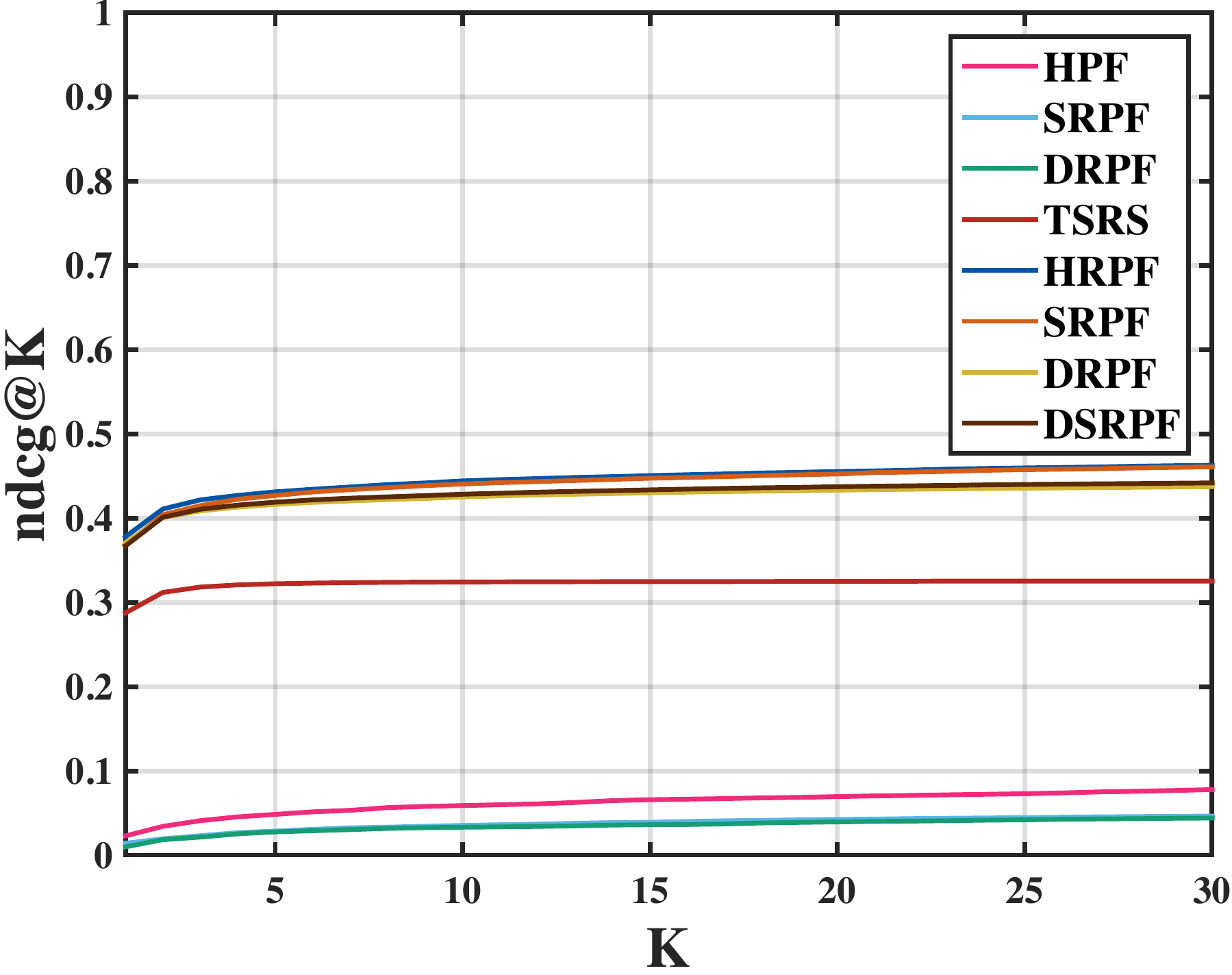}%
\label{fig:LastFM_overK}}
\vfil
\makebox[10pt]{\raisebox{40pt}{\rotatebox[origin=c]{90}{\bf Tianchi}}}
\subfloat[]{\includegraphics[width=1.6in]{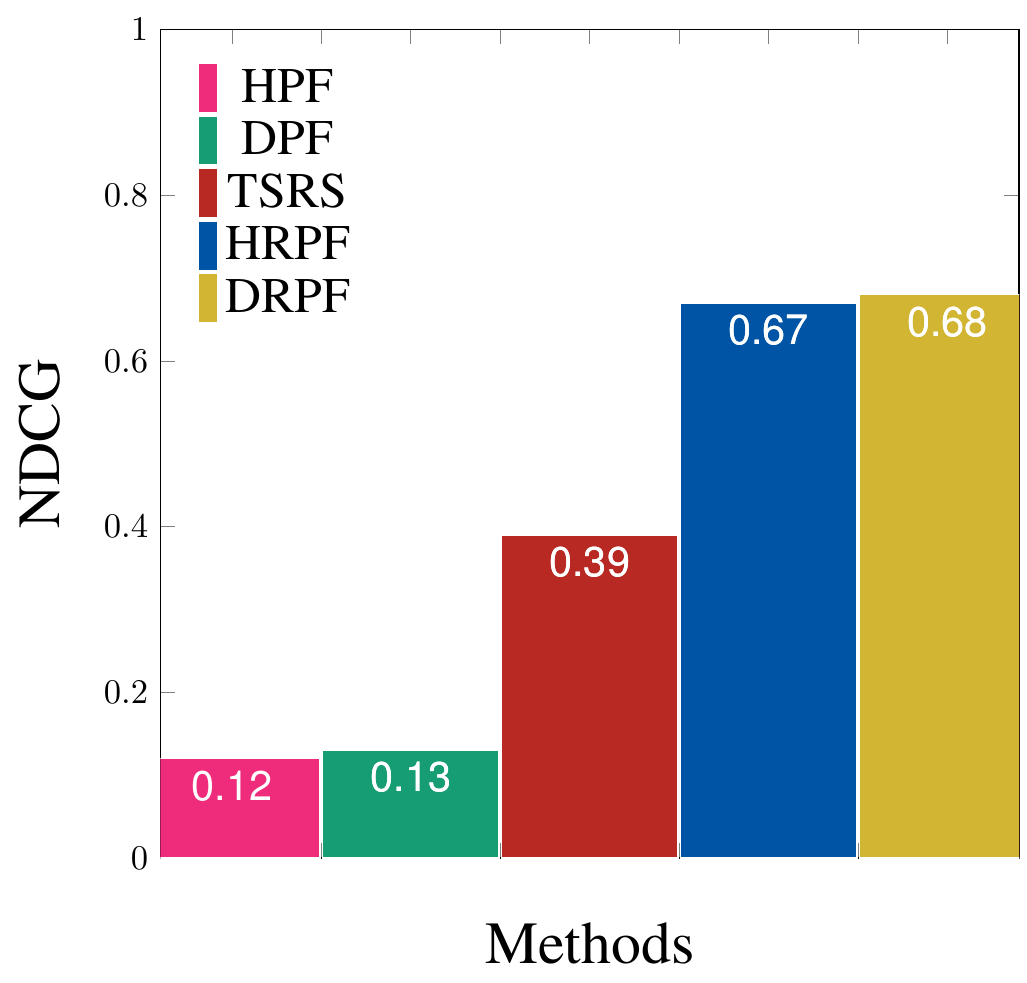}%
\label{fig:NDCG_Tianchi}}
\hspace{0.01in}%
\subfloat[]{\includegraphics[width=1.6in]{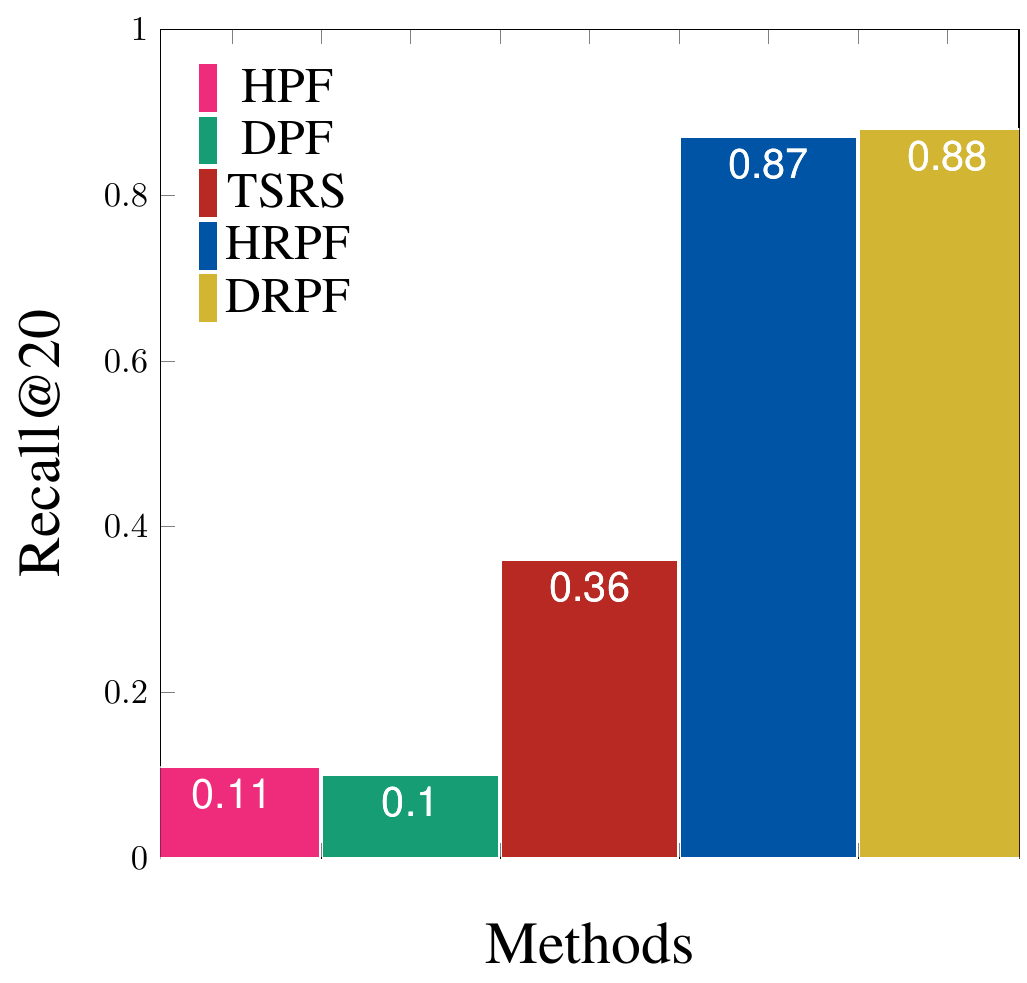}%
\label{fig:RecallAt20_Tianchi}}
\hspace{0.1in}%
\subfloat[]{\includegraphics[height=1.5in,width=1.6in]{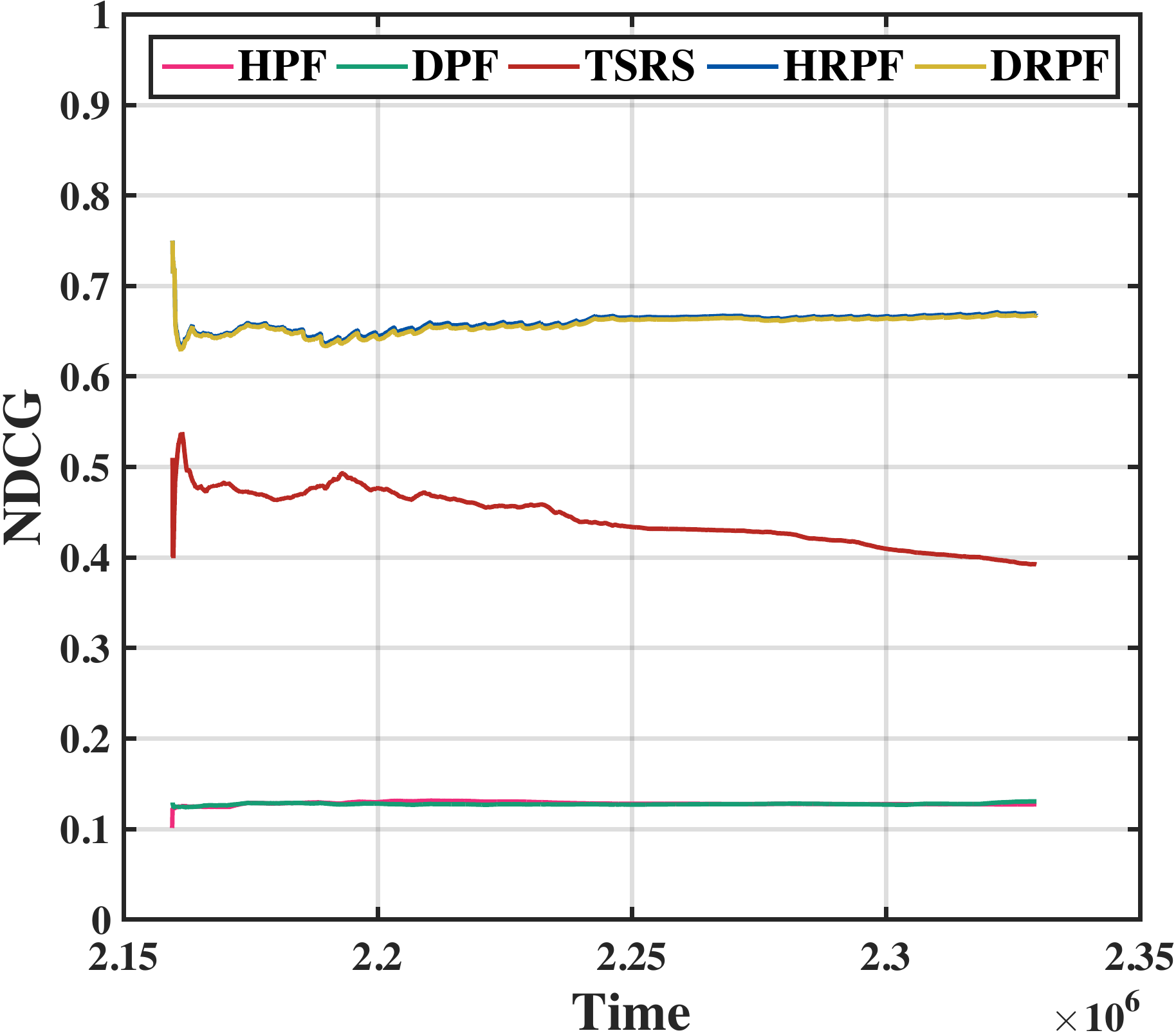}%
\label{fig:Tianchi_over_time}}
\hspace{0.1in}%
\subfloat[]{\includegraphics[height=1.5in,width=1.6in]{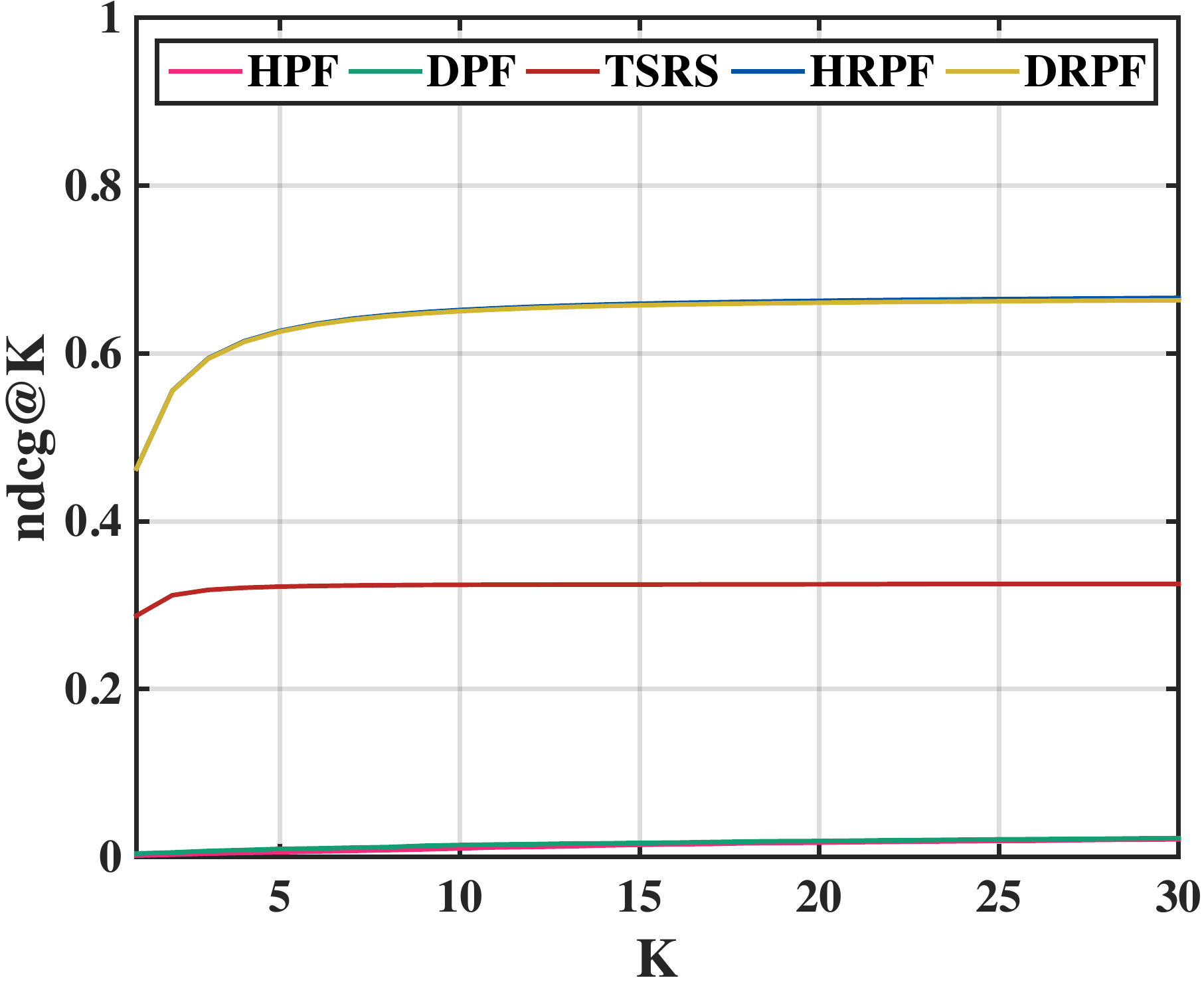}%
\label{fig:Tianchi_overK}}
\caption{ Performance of different methods in item recommendations. Top row presents the \texttt{NDCG@20} and \texttt{Recall@20} for \textbf{Last.fm} dataset, and the bottom row presents the results for \textbf{Tianchi} dataset.
}
\label{fig:real_errors}
\end{figure*}
\subsection{Competitor baselines}
To show the performance of RPF in capturing the dynamic preferences and the recurrent consumption behavior we compare it with the recent state-of-the-art PF-based methods. Notably, we compared with \textbf{DPF}~\cite{DPF}, which captures the dynamic user preferences, \textbf{SPF}~\cite{SPF}, that 
 accounts for the social aspects of product adoption, and \textbf{HPF}~\cite{HPF}, which utilizes a hierarchical structure to better model diverse user preferences. We also selected the Time sensitive recommendation system \textbf{TSRS}~\cite{TimeSensitive}, which is a low rank method based on Hawkes process and models user-item interactions when the users are independent. 


Also, in order to understand RPF better we evaluated different variants of our proposed framework. They include
\textbf{SRPF}, which is a variant of the RPF, which considers the influence-based preference among users;  \textbf{DRPF}, that captures the dynamically changing latent preference in the RPF; and  \textbf{DSRPF}, which is a version of RPF that enables the peer influence on the social network affects dynamic user preferences.	

\subsection{Synthetic Data}
We first study the effectiveness of the proposed inference algorithm on synthetic data to make sure the model is learnable given a reasonable amount of observations.

\textbf{Experimental Setup.}
Our synthetic recommendation task consists of 1000 users and 1000 items. We generated a random network between users with an average degree 50. The parameters of RPF are sampled from the associated distributions introduced in the  section~\ref{sec:proposed}.  Finally, we generated about 1M events using the Ogata's thinning algorithm \cite{ogata1981lewis}. We repeated each experiment  10 times and the results are reported by taking the average over these runs. It is worth mentioning that the events are generated using the proposed DSRPF model.

\textbf{Results.}
Figure \ref{fig:synt_performance} presents the  performance of proposed RPF framework over synthetic data.
We reported the Mean Absolute Error (MAE) between the true parameters of the model and the estimated ones. Figure \ref{fig:synt_performance}a and \ref{fig:synt_performance}b shows the performance of the proposed inference procedure. Furthermore,  the increased number of iterations results in a better fit. More especially, Figure \ref{fig:MAEOverIterations_Synth} shows that the proposed method only requires around 300 iterations to converge which shows the fast convergence of the inference algorithm. 
Figure \ref{fig:MAEOverEvents_Synth} presents the MAE over different number of events. As it is expected, with the increase in the number of events the performance of proposed method improves as well. Furthermore, it verifies that the proposed inference algorithm only requires a modest number of events to achieve a good performance. 
Additionally, we investigate how well the temporal patterns in data is captured by the proposed algorithm. 
According to the time-change theorem \cite{PointProcessesTheory}, given all   successive event times of a particular point process with intensity function $\lambda(t)$ , the set of intensity integrals $\int_{t_i}^{t_{i+1}}\lambda(t)dt$ should conform to the unit-rate exponential distribution, if the samples are truly generated from the process $\lambda(t)$.
Figure \ref{fig:QQPlot_Synth} demonstrates the quantiles of intensity integrals computed using the learned intensities. The conformity of the empirical quintiles and the true ones suggests that the model captures the temporal dynamics very well.   Last but not least, we investigate the scalability of algorithm. Figure \ref{fig:Scalability_Synth} shows that with the increase in the number of events the inference algorithm scales almost linearly.

\begin{figure}[!t]
\vspace{-7mm}
\centering
{\includegraphics[width=2.5in]{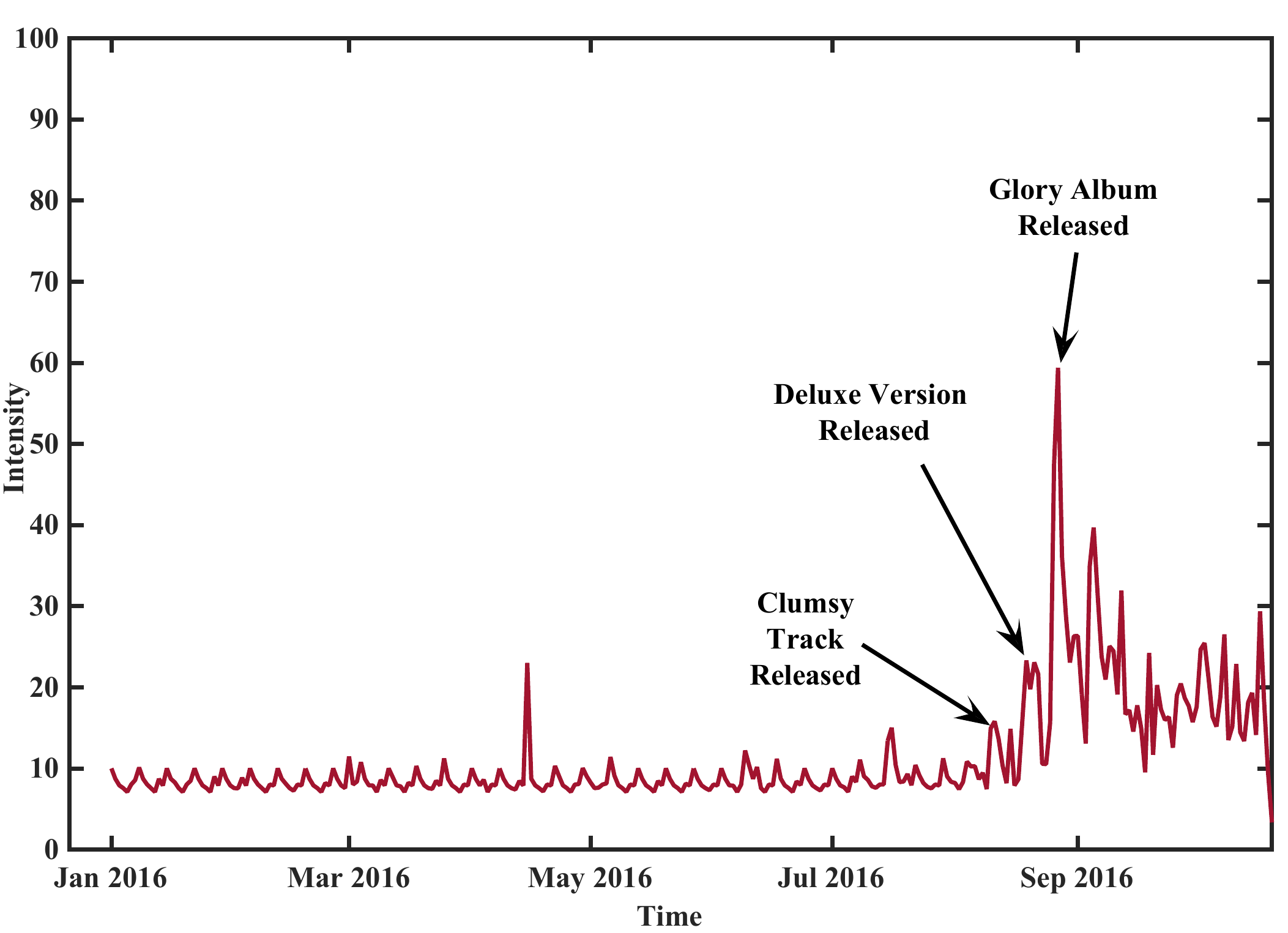}}
\caption{ The total learned intensity of users for listening to Britney Spears songs over time. 
}
\label{fig:qualitative_britney}
\vspace{-7mm}
\end{figure}

\subsection{Real Data}

We also evaluate the performance of RPF on two real datasets from different domains; \emph{Last.fm} and \emph{Alibaba Mobile Commerce}. Both datasets contain timestamped actions which makes them appropriate benchmark for comparing the proposed method to the other state-of-the-art methods.

\textbf{Last.fm.}
This dataset contains the music listening events between 1200 users and 1000 artists. There are around 480000 events in total which spans a period of six months~\cite{Celma:Springer2010}. 

\textbf{Tianchi Mobile Data.} It contains the user interactions with items in Alibaba's mobile M-Commerce platform~\cite{alibabaDataSet}. The dataset includes four behavior types: click, collect, add-to-cart, and payment. We only considered  the click events and used the item categories as the recommendation targets. Our data contains  roughly 1000 users , 2100 items, and the total 1200000 events.



\textbf{Qualitative Analysis.}
We first explored the fitted model to confirm that the model can capture different patterns in data. Figure \ref{fig:qualitative_britney} represents the summation of the learned intensity of users listening to the Britney Spears thorough time in the Last.fm dataset. We see a sudden increase in the learned intensity at 26th august 2016, which is the time her new album, \emph{Glory}, was released.  It is interesting that  there are two other jumps in the learned intensity before this sudden increase at  August 11th and August 21st , which are when the track  \emph{Clumsy}  and \emph{Deluxe} version of the album  was released.  This is how RPF framework can capture different temporal dynamics in item consumptions. 

\begin{figure}[!t]
\vspace{-7mm}
\centering
\subfloat[]{\includegraphics[width=1.5in]{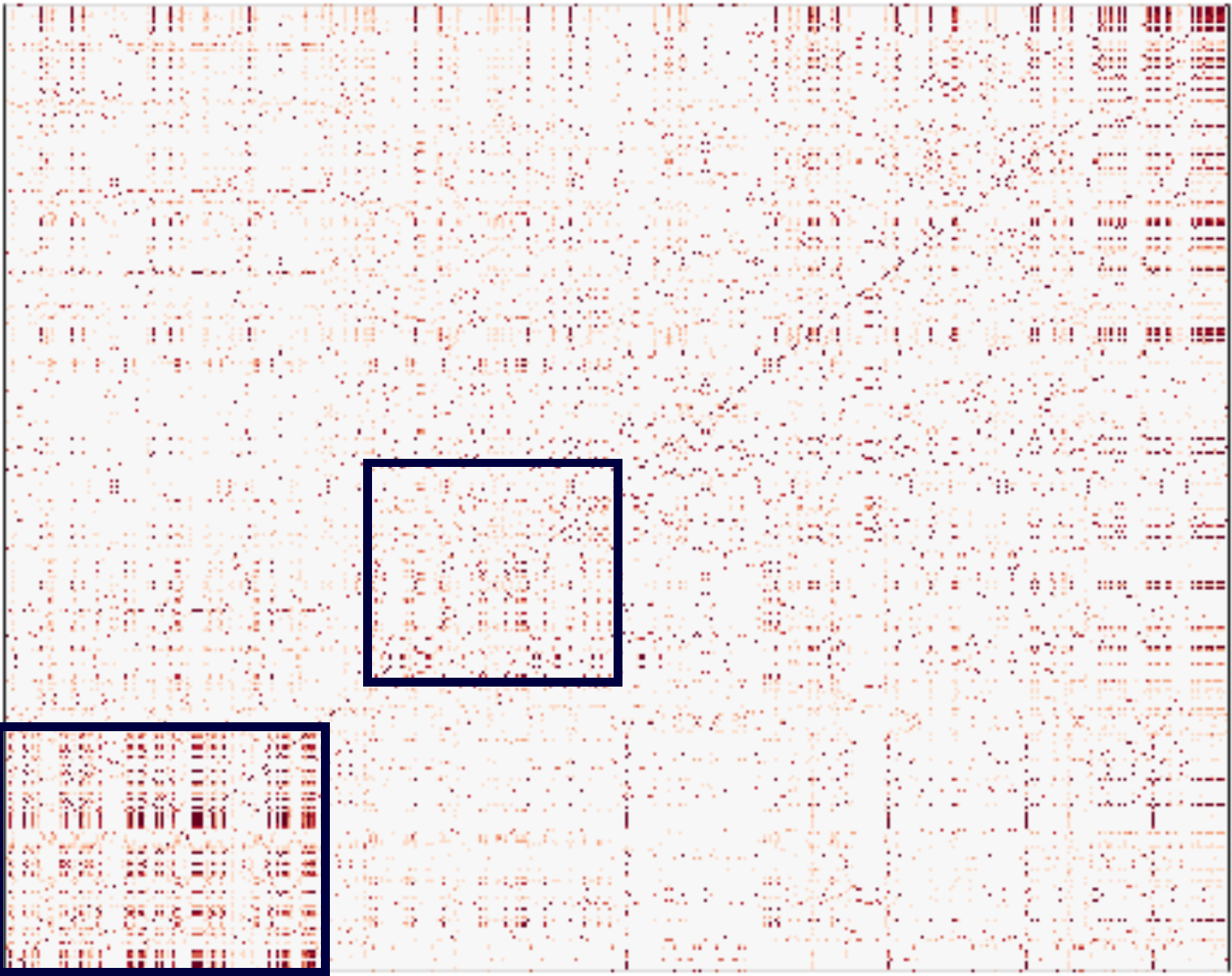}%
\label{fig:TheoreticalSimilarity_LastFM}}
\hspace{0.1in}%
\subfloat[]{\includegraphics[width=1.5in]{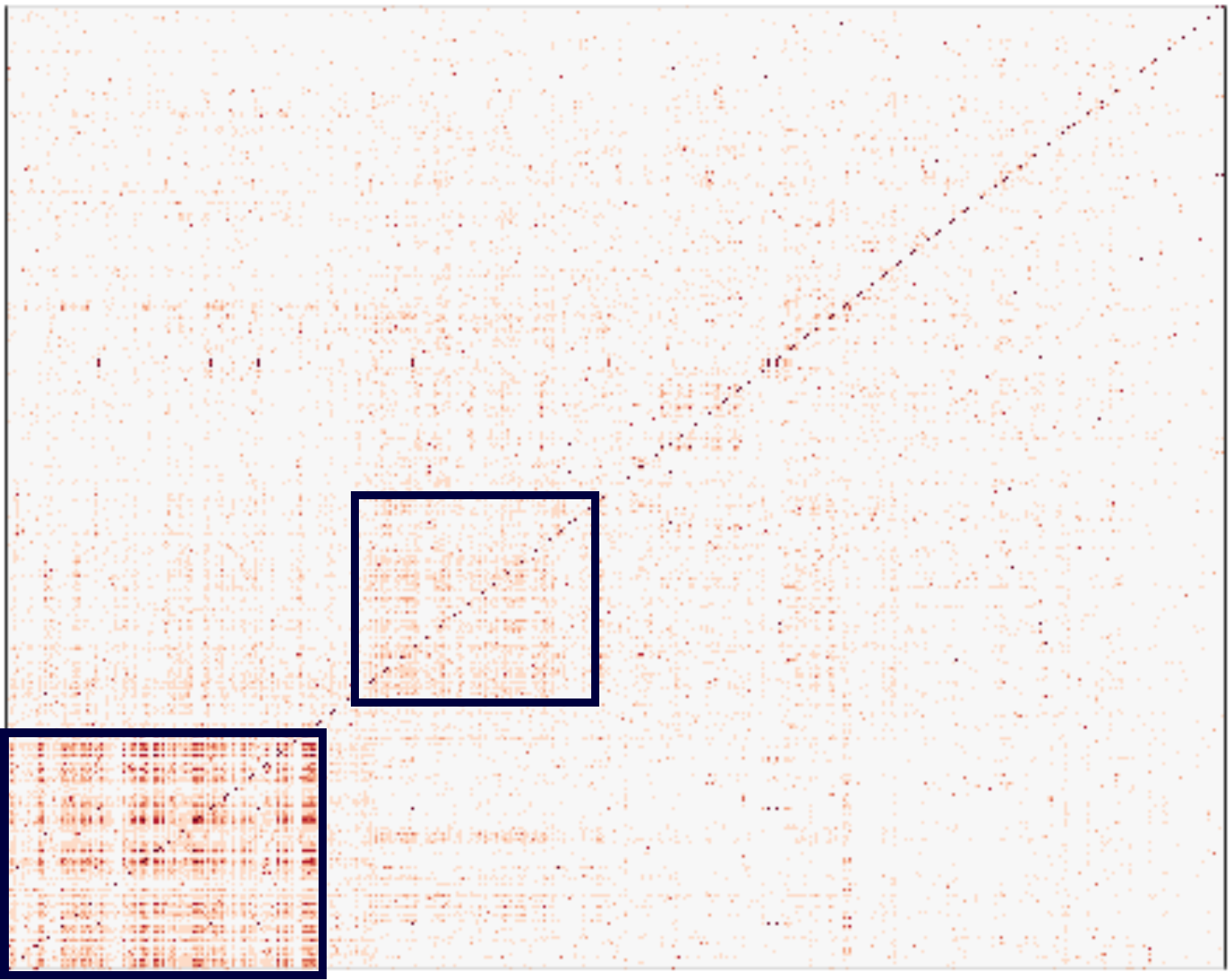}%
\label{fig:EmpiricalSimilarity_LastFM}}
\caption{(a) Learned similarity between user interests, and (b) The empirical Jacard similarity between user listening behaviors over Last.fm dataset.
}
\label{fig:qualitative_simliarity}
\vspace{-7mm}
\end{figure}

Next, we investigated how much the proposed method can capture the similarity between users' preferences. To this end, we created the similarity matrix using the learned user preferences $\theta_u$. Each $a_{uv}$ entry of the defined similarity matrix is equal to $\theta_u^\top \theta_v$ which indicates that how much the latent preferences of users $u$ and $v$ are close.  This is roughly proportional to their shared items of interest.
We also created an empirical similarity matrix, using the Jaccard similarity  between all pair of users based on the  size of shared consumed items.
Figures \ref{fig:qualitative_simliarity}a and \ref{fig:qualitative_simliarity}b presents the learned similarity matrix and the empirical one for the Last.fm dataset. Interestingly, these two matrices conform by a reasonable amount. 
We also highlighted the two  blocks of highly similar users in the matrices with squares. As it can be seen, there is a strong correlation between the two matrices, and the blocks match each other. It basically shows the proposed model can efficiently learn user preferences and can capture their similarities.


\textbf{Item Prediction.}
Our first quantitative task is  predicting the test items that each user will interact with. 
We generate top $k$ recommendations for each user, using  items with the highest probability under each method. The methods are evaluated based on $\text{NDCG@k}$ and $\text{Recall@k}$ metrics:
\begin{align}
	\text{Recall@k} = \frac{1}{N}\sum_{n=1}^{N}\mathcal{I}(rank(p_n)<k)\\
	\text{NDCG@k} = \frac{1}{N}\sum_{n=1}^{N}\frac{\mathcal{I}\left(rank(p_n)<k\right)}{\log_{2}{1+rank(p_n)}}
\end{align}
where $N$ is the number of test data and $\mathcal{I}$ is the indicator function. $\text{Recall@K}$ shows the percentage of items that are ranked in the first $k$ . $\text{NDCG@k}$ is a weighted version of $\text{Recall@K}$ which 
The interested reader is  referred to \cite{DPF} for their detailed properties.

Figures \ref{fig:NDCG_LastFM},\ref{fig:RecallAt20_LastFM},\ref{fig:NDCG_Tianchi}, and \ref{fig:RecallAt20_Tianchi} demonstrate the \texttt{NDCG@k} and \texttt{Recall@k} of different methods over Last.fm and Tianchi datasets, respectively.
It is noteworthy that since there is no social network in Tianchi data, the result of social based methods (SPF, SRPF, and DSRPF) are not reported on it.
 For item prediction, RPF-based methods perform significantly better than competitors. We believe this is mostly due to the fact that it can handle temporal nature of the data inherently. Furthermore,
 TSRS performs marginally better than PF-based methods which is also as expected. It treats time in a proper way. Interestingly, TSRS margin with respect to PF-based methods is higher in Last.fm dataset compared to Tianchi.
 To explain this observation note that  Last.fm's items are the artists, and users play the songs of an artist sequentially in the sessions. Hence, the history of previous interactions play an important role in determining the next action.  TSRS itself has a more emphasis on previous user interactions than the intrinsic user preferences.
 On the other hand, the Tianchi dataset contains the clicks of users over different items. Therefore, capturing the preferences of users is more central than paying attention to the timings. Therefore, the margin of TSRS and PF-based methods is smaller in this dataset.

We also evaluated the impact of size of recommendation list ($k$) and  the size of test data on the performance of different methods. Figure \ref{fig:LastFM_over_time} and \ref{fig:Tianchi_over_time} show the \texttt{NDCG@20} for different methods over different sizes of test data. As it can be seen, with the increase in the number of test events the performance of different methods degrades, however, RPF-based methods perform marginally better than competitors in both data sets and the relative order of different methods are preserved with the increase in data. 

Figures \ref{fig:LastFM_overK} and \ref{fig:Tianchi_overK} show the \texttt{NDCG@k} for different methods over different values of $k$. All methods exhibit an early increase in \texttt{NDCG@k} with increase of $k$. Moreover, RPF-based methods consistently being the best ones.

 In a nutshell, TSRS pays more attention to the impact of previous events of the user, and less attention is paid to the exact modeling of intrinsic user preferences. It uses constraint optimizations to implicitly consider the impact of different items on each other, while in the RPF-based methods the user preferences are modeled using explicit random variables and are inferred through matrix factorization. Hence the RPF-based methods can better capture both the intrinsic user preferences and the impact of previous events. Among the RPF-based methods, DRPF has many parameters and needs more data to be trained. Over the Last.fm data the  number of events per user are significantly fewer than the Tianchi data. Therefore, the DRPF performs better over Tianchi data. 
 

%
%

\begin{figure}[!t]
\vspace{-5mm}
\centering
\makebox[10pt]{\raisebox{40pt}{\rotatebox[origin=c]{90}{\bf Last.fm}}}
\subfloat[]{\includegraphics[height=1.3in,width=1.5in]{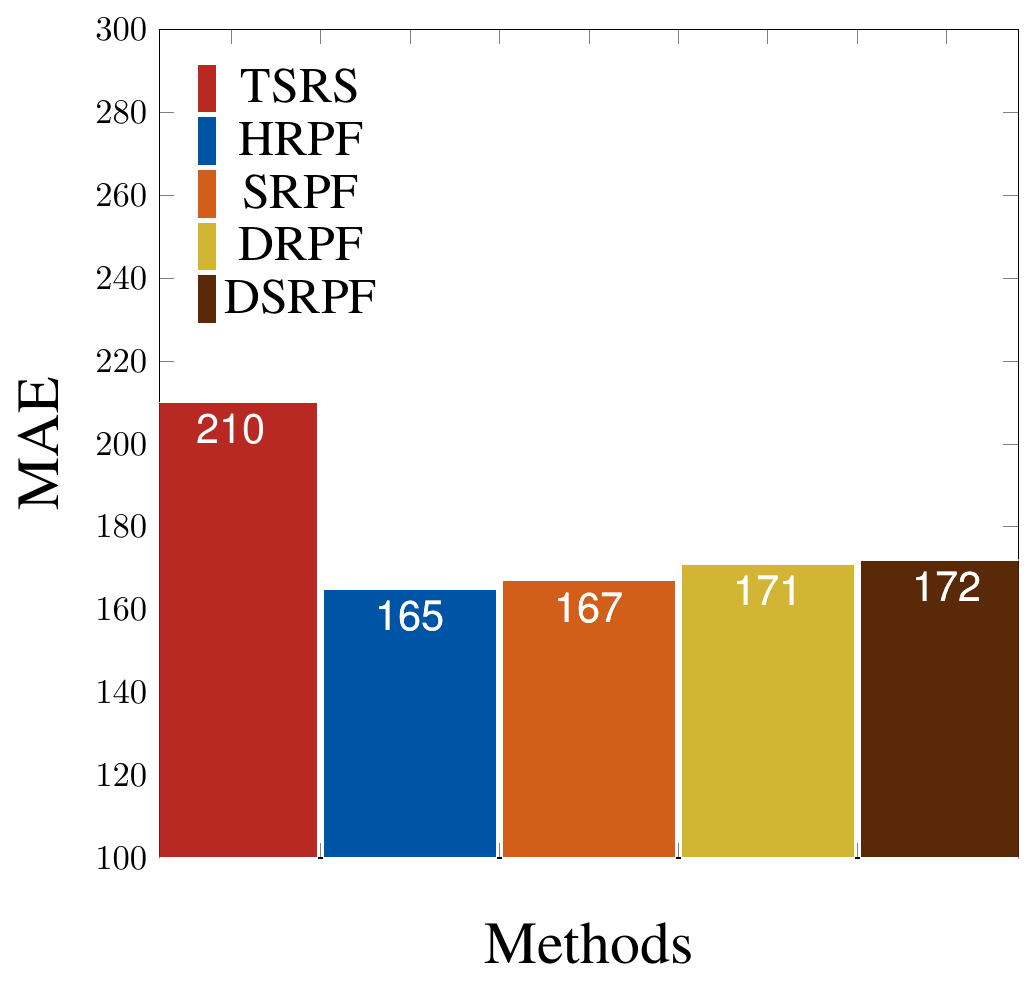}%
\label{fig:synt_mre_time_overal}}
\hspace{0.01in}%
\subfloat[]{\includegraphics[height=1.4in,width=1.6in]{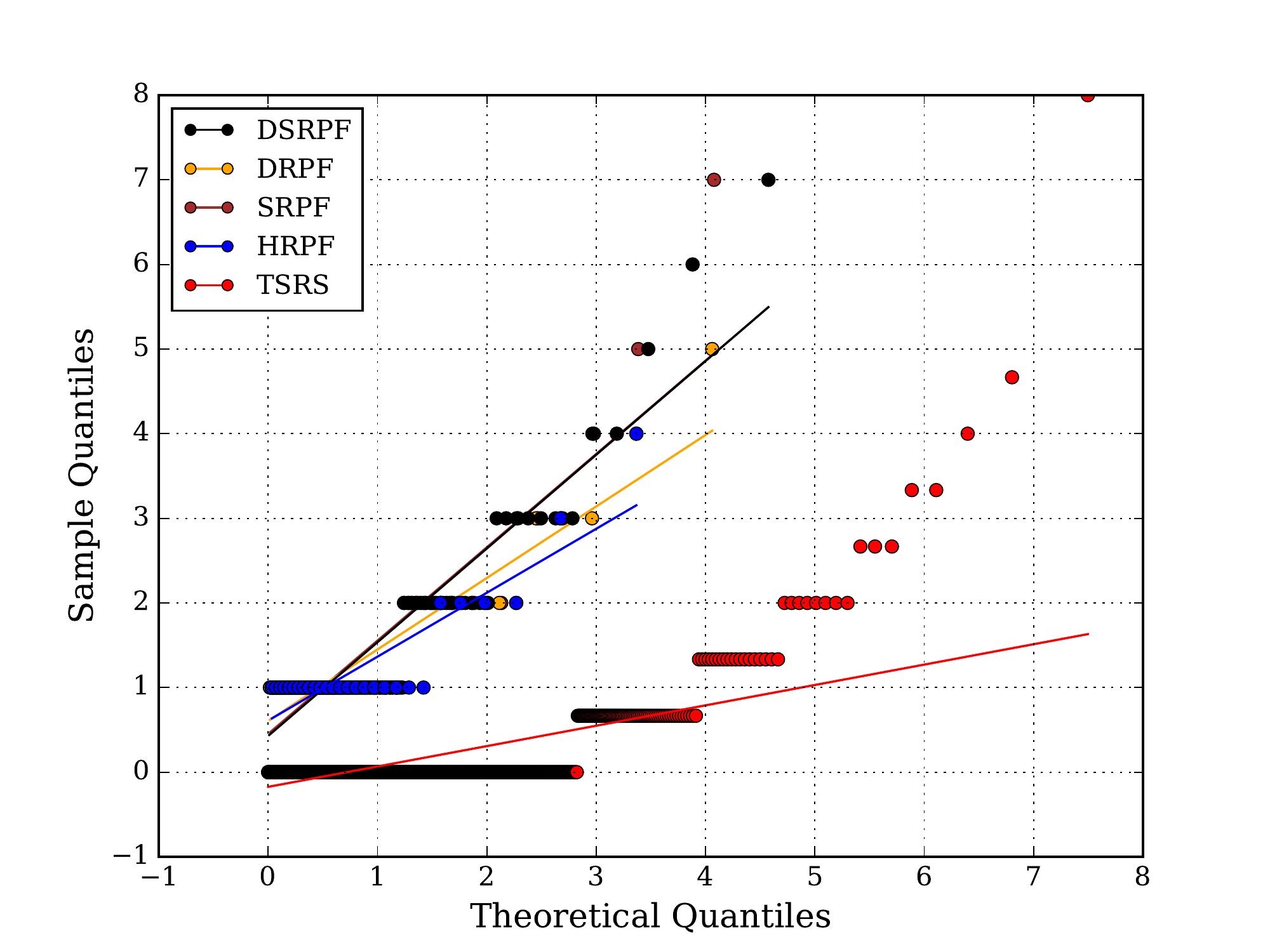}%
\label{fig:synt_rank_time_overal}}
\vfil
\makebox[10pt]{\raisebox{40pt}{\rotatebox[origin=c]{90}{\bf Tianchi}}}
\subfloat[]{\includegraphics[height=1.3in,width=1.5in]{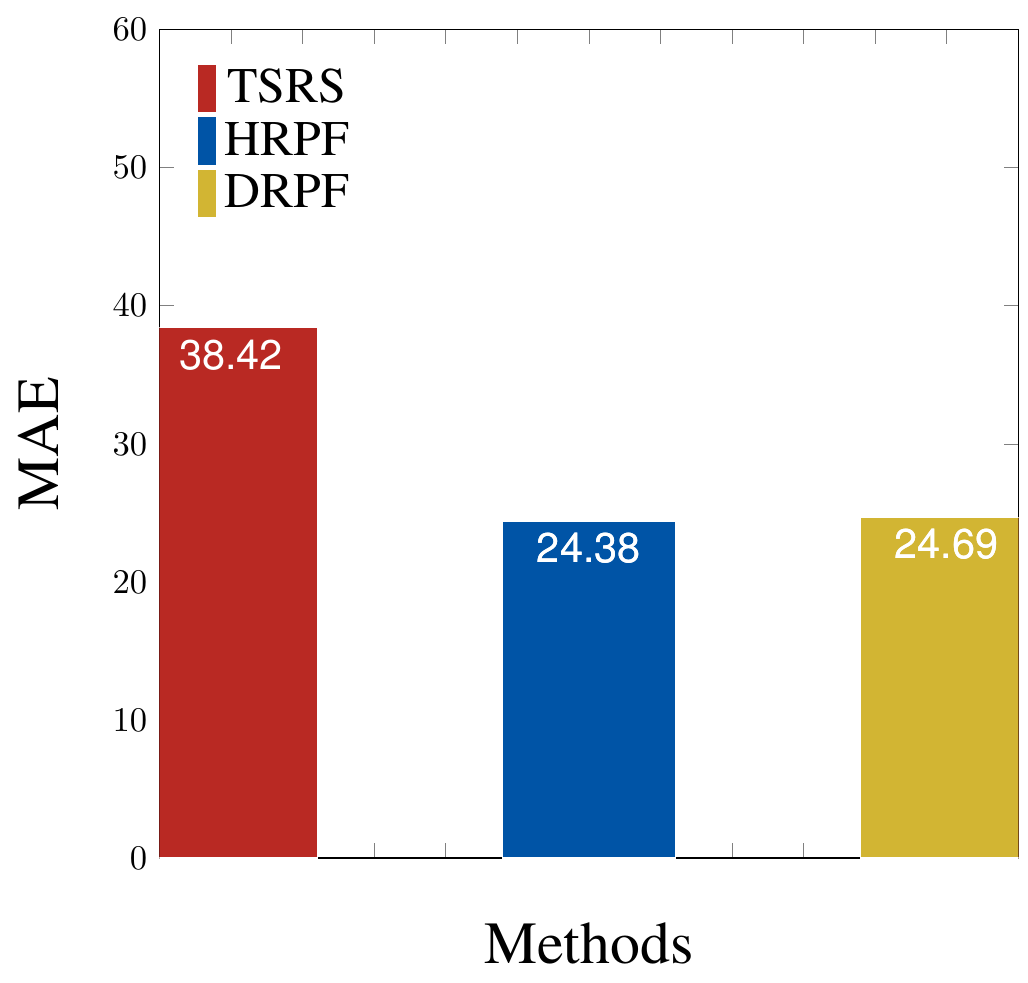}%
\label{fig:synt_mre_topic_overal}}
\hspace{0.01in}%
\subfloat[]{\includegraphics[height=1.4in,width=1.6in]{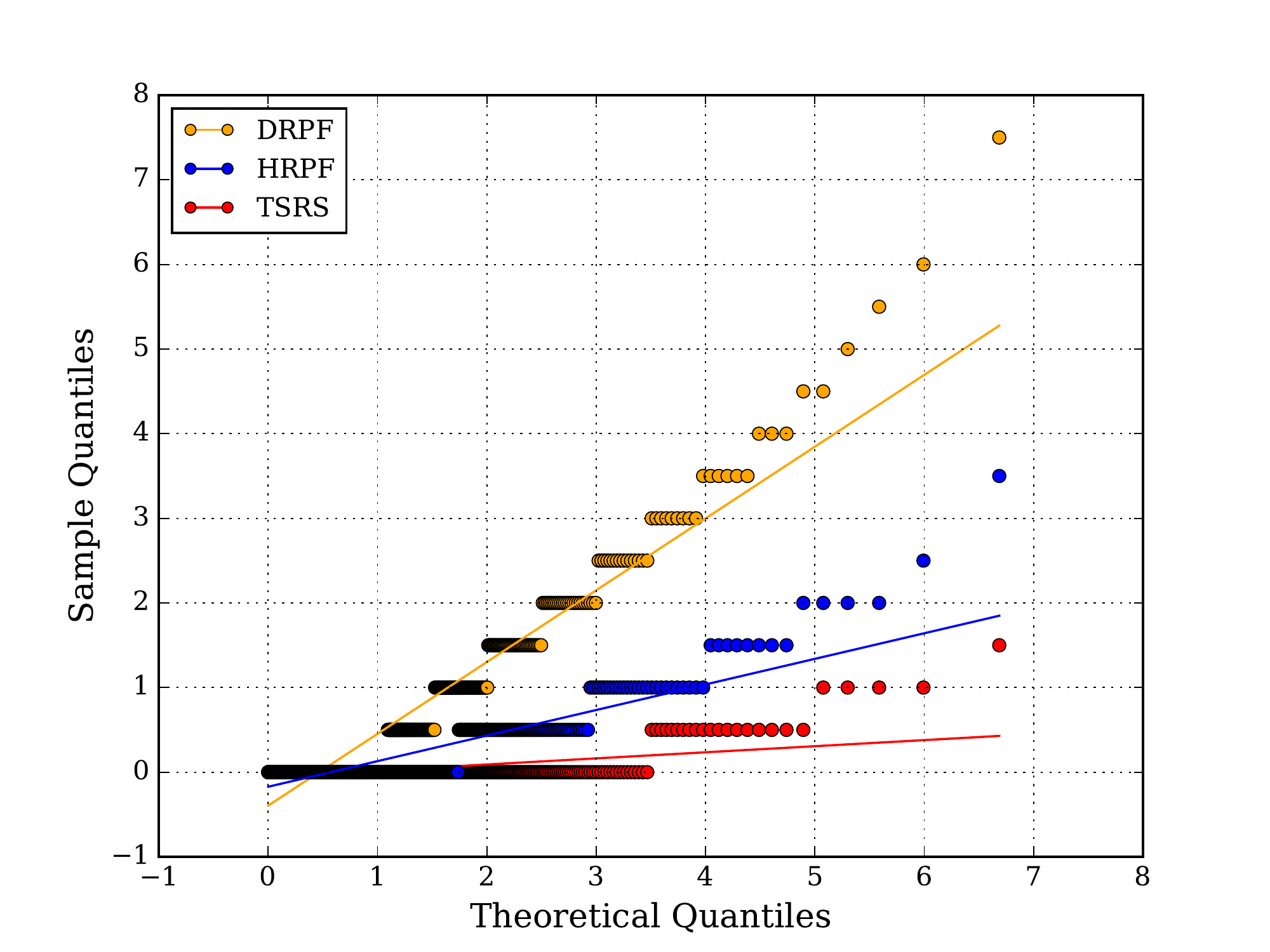}%
\label{fig:synt_rank_topic_overal}}
\vspace{-3mm}
\caption{Performance of different methods on returning-time prediction. 
(a), (c) MAE of different methods in predicting the returning-time of users for Last.fm and Tianchi datasets. (b), (d) Quantile plot of  different methods for Last.fm and Tianchi datasets.
}
\label{fig:time_prediction_results_real}
\vspace{-3mm}
\end{figure}

\textbf{Returning Time Prediction.}
Finally, we evaluated the performance of different methods in predicting the time when a user will return to the system. To this end, we predict the next time a user will do an action and compute the MAE between the estimated time and the ground-truth time. We report the average results over all users in Figure \ref{fig:time_prediction_results_real}. 
Figures \ref{fig:time_prediction_results_real}a and \ref{fig:time_prediction_results_real}c show that all the RPF-based methods perform better than the TSRS. This is due to the fact that the proposed RPF-based methods model the dynamics of user interests and consider the peer influence of the neighboring users consuming a product.
Hence, they are more expressive than the TSRS  which assumes that the user-item interactions are independent.

We also plotted the Quantile plot for different methods. To this end, we compare the theoretical quantiles from the exponential distribution with the ones that each model has learnt from real-world data. The closer the slope is to one, the better a model matches the event patterns.
Figures \ref{fig:time_prediction_results_real}b and  \ref{fig:time_prediction_results_real}d show that the RPF-based models fit the observed event sequence better than the TSRS. Over the Last.fm dataset,  DSRPF that utilizes both the peer influence and the dynamic user preferences, better fits the real data. On the Tianchi recommendation task, the DRPF better fits the real data. This is due to the large number of events per user-item on this dataset, which helps the DRPF to capture the true dynamics of real data.

\section{Conclusions}
We present a novel framework, \emph{Recurrent Poison Factorization} (RPF), for building recommendation systems from implicit feedback data. 
RPF extends the Poisson factorization (PF) methods  by modeling the aggregated count data with a Poisson process, therefore, enabling the model to account for recurrent activities and answering time-sensitive queries. 
Moreover, The proposed variational inference is able to scale to large datasets using implicit information.
RPF can handle the change of users and items specification through time (DRPF), can consider socially-influenced user preferences (SRPF), and is able to capture the heterogeneity amongst users and items (HRPF). 
Experiments on synthetic data and two real world datasets demonstrate superiority of the proposed framework over several state-of-the-art methods on item prediction and returning time prediction -- thanks to its temporal capabilities. 
Furthermore, in terms of interpretability, our model is able to discover interesting patterns such as peer influence between users, trending products, and their temporal properties. 

For future work, we would like to incorporate nonparametric models into RPF in order to add more flexibility on capturing  dynamic preferences and peer influences.  Another interesting venue for future work is utilizing generative deep neural networks to model the intensity functions. 
\bibliographystyle{ACM-Reference-Format}
\bibliography{RPF_References} 

\end{document}